\documentclass[twoside,twocolumn,9pt]{article}
\usepackage{extsizes}
\usepackage[super,sort&compress,comma]{natbib} 
\usepackage[version=3]{mhchem}
\usepackage[left=1.5cm, right=1.5cm, top=1.785cm, bottom=2.0cm]{geometry}
\usepackage{balance}
\usepackage{times,mathptmx}
\usepackage{sectsty}
\usepackage{graphicx} 
\usepackage{lastpage}
\usepackage[format=plain,justification=justified,singlelinecheck=false,font={stretch=1.125,small,sf},labelfont=bf,labelsep=space]{caption}
\usepackage{float}
\usepackage{fancyhdr}
\usepackage{fnpos}
\usepackage[english]{babel}
\usepackage{array}
\usepackage{droidsans}
\usepackage{charter}
\usepackage[T1]{fontenc}
\usepackage[usenames,dvipsnames]{xcolor}
\usepackage{setspace}
\usepackage[compact]{titlesec}

\usepackage{siunitx} 
\usepackage{placeins} 
\usepackage{subcaption} 
\usepackage{dcolumn}
\usepackage{multirow}
\usepackage{booktabs}
\usepackage{url}

\usepackage{epstopdf}

\definecolor{cream}{RGB}{222,217,201}

\begin{document}

\pagestyle{fancy}
\thispagestyle{plain}
\fancypagestyle{plain}{

\fancyhead[C]{}
\fancyhead[L]{}
\fancyhead[R]{}
\renewcommand{\headrulewidth}{0pt}
}

\makeFNbottom
\makeatletter
\renewcommand\LARGE{\@setfontsize\LARGE{15pt}{17}}
\renewcommand\Large{\@setfontsize\Large{12pt}{14}}
\renewcommand\large{\@setfontsize\large{10pt}{12}}
\renewcommand\footnotesize{\@setfontsize\footnotesize{7pt}{10}}
\makeatother

\renewcommand{\thefootnote}{\fnsymbol{footnote}}
\renewcommand\footnoterule{\vspace*{1pt}%
\color{cream}\hrule width 3.5in height 0.4pt \color{black}\vspace*{5pt}} 
\setcounter{secnumdepth}{5}

\makeatletter 
\renewcommand\@biblabel[1]{#1}            
\renewcommand\@makefntext[1]%
{\noindent\makebox[0pt][r]{\@thefnmark\,}#1}
\makeatother 
\renewcommand{\figurename}{\small{Fig.}~}
\sectionfont{\sffamily\Large}
\subsectionfont{\normalsize}
\subsubsectionfont{\bf}
\setstretch{1.125} 
\setlength{\skip\footins}{0.8cm}
\setlength{\footnotesep}{0.25cm}
\setlength{\jot}{10pt}
\titlespacing*{\section}{0pt}{4pt}{4pt}
\titlespacing*{\subsection}{0pt}{15pt}{1pt}

\fancyfoot{}
\fancyfoot[LO,RE]{\vspace{-7.1pt}}
\fancyfoot[CO]{\vspace{-7.1pt}\hspace{13.2cm}}
\fancyfoot[CE]{\vspace{-7.2pt}\hspace{-14.2cm}}
\fancyfoot[RO]{\footnotesize{\sffamily{ ~\textbar \hspace{2pt}\thepage}}}
\fancyfoot[LE]{\footnotesize{\sffamily{\thepage~\textbar\hspace{3.45cm}}}}
\fancyhead{}
\renewcommand{\headrulewidth}{0pt} 
\renewcommand{\footrulewidth}{0pt}
\setlength{\arrayrulewidth}{1pt}
\setlength{\columnsep}{6.5mm}
\setlength\bibsep{1pt}

\makeatletter 
\newlength{\figrulesep} 
\setlength{\figrulesep}{0.5\textfloatsep} 

\newcommand{\topfigrule}{\vspace*{-1pt}%
\noindent{\color{cream}\rule[-\figrulesep]{\columnwidth}{1.5pt}} }

\newcommand{\botfigrule}{\vspace*{-2pt}%
\noindent{\color{cream}\rule[\figrulesep]{\columnwidth}{1.5pt}} }

\newcommand{\dblfigrule}{\vspace*{-1pt}%
\noindent{\color{cream}\rule[-\figrulesep]{\textwidth}{1.5pt}} }

\makeatother

\twocolumn[
  \begin{@twocolumnfalse}
\vspace{1cm}
\noindent\LARGE{\textbf{First-principles analysis of the spectroscopic limited maximum efficiency of photovoltaic absorber layers for CuAu-like chalcogenides and silicon}}
\sffamily
\begin{tabular}{m{1cm} p{15.5cm} m{1cm}}

\vspace{0.3cm} & \vspace{0.3cm} & \vspace{0.3cm} \\

 & \centering\noindent\large{Marnik Bercx$^{\ast}$\textit{$^{a}$}, Nasrin Sarmadian\textit{$^{a}$}, Rolando Saniz\textit{$^{a}$}, Bart Partoens\textit{$^{a}$} and Dirk Lamoen\textit{$^{a}$}} & \\
 \vspace{0.3cm} & \vspace{0.3cm} & \vspace{0.3cm} \\
 
& \noindent\normalsize{Chalcopyrite semiconductors are of considerable interest for application as absorber layers in thin-film photovoltaic cells. When growing films of these compounds, however, they are often found to contain CuAu-like domains, a metastable phase of chalcopyrite. It has been reported that for CuInS$_2$, the presence of the CuAu-like phase improves the short circuit current of the chalcopyrite-based photovoltaic cell. We investigate the thermodynamic stability of both phases for a selected list of I-III-VI$_2$ materials using a first-principles density functional theory approach. For the \mbox{CuIn-VI$_2$} compounds, the difference in formation energy between the chalcopyrite and CuAu-like phase is found to be close to 2~\si{\milli\electronvolt}/atom, indicating a high likelihood of the presence of CuAu-like domains. Next, we calculate the Spectroscopic Limited Maximum Efficiency (SLME) of the CuAu-like phase and compare the results with those of the corresponding chalcopyrite phase. We identify several candidates with a high efficiency, such as CuAu-like CuInS$_2$, for which we obtain an SLME of 29\% at a thickness of 500~\si{\nano\meter}. We observe that the SLME can have values above the Shockley-Queisser (SQ) limit, and show that this can occur because the SQ limit assumes the absorptivity to be a step function, thus overestimating the radiative recombination in the detailed balance approach. This means that it is possible to find higher theoretical efficiencies within this framework simply by calculating the $J$-$V$ characteristic with an absorption spectrum. Finally, we expand our SLME analysis to indirect band gap absorbers by studying silicon, and find that the SLME quickly overestimates the reverse saturation current of indirect band gap materials, drastically lowering their calculated efficiency.} \\ &

\end{tabular}

 \end{@twocolumnfalse} \vspace{1cm}

  ]

\renewcommand*\rmdefault{bch}\normalfont\upshape
\rmfamily
\section*{}
\vspace{-1cm}


\footnotetext{\textit{$^{a}$~EMAT \& CMT groups, Department of Physics, University of Antwerp, Belgium.}}
\footnotetext{\textit{$^{*}$~Campus Groenenborger, Groenenborgerlaan 171, 2020 Antwerp, Belgium. Tel:~+3232653572; E-mail:~marnik.bercx@uantwerpen.be}}

\footnotetext{\dag~Supplementary Information can be found at the end of the document.}



\section{\label{sec:intro}Introduction}

The conventional search for potential absorber materials in photovoltaic devices is expensive and time consuming. Inverse design methods have the power to screen materials relatively quickly, providing valuable information that allows experimental work to focus on promising compounds~\cite{Jain2013}. In order to accurately screen materials, however, a proper selection metric is required. Traditionally, the Shockley-Queisser (SQ) limit~\cite{Shockley1961} has been used as a theoretical gauge of the potential efficiency of absorbers. The Spectroscopic Limited Maximum Efficiency~\cite{Yu2012} (SLME) goes beyond the SQ limit by including the absorption spectrum and film thickness in the determination of the efficiency. Since its conception, the SLME has been successfully applied to perovskites~\cite{Yin2014,Yin2015,Yin2015b,Meng2016}, chalcogenides~\cite{Hong2016,Nasrin2016}, direct band gap silicon crystals~\cite{Lee2014,Oh2015} and other materials~\cite{Yu2012b,Yokoyama2013,Heo2014,Huang2015}.

Ternary I-III-VI$_2$ semiconductors, such as the well known Cu(In,Ga)(S,Se)$_2$ compounds, are commonly used as absorber materials to produce highly flexible and lightweight solar cells. The high absorption coefficient of these compounds allows for cost-efficient absorber layers that are particularly suited for deposition on flexible substrates~\cite{Reinhard2013}. Laboratory values for the efficiency of CuIn(S,Se)$_2$ thin film solar cells have recently reached a record value of 22.3\% \cite{Solarfrontier}. Furthermore, CuIn(S,Se)$_2$ is also considered a suitable material for the top cell in tandem structures~\cite{Cheek2013} and quantum dot based luminescent solar concentrators~\cite{Hu2015}. The rapid succession of new record efficiencies indicates that there is still room for improvement in these applications.

\begin{figure}[htbp] 
\setlength{\captionmargin}{10pt}
\centering
\begin{subfigure}{0.24\textwidth}
\centering
\includegraphics[width=0.8\linewidth]{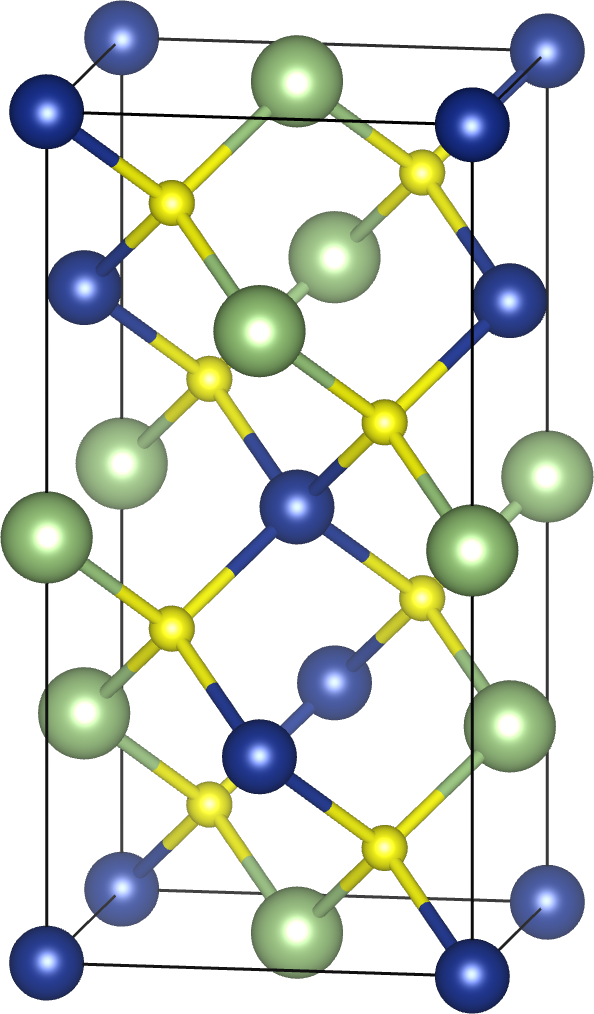}
\caption{}
\end{subfigure}%
\begin{subfigure}{0.24\textwidth}
\centering
\includegraphics[width=0.8\linewidth]{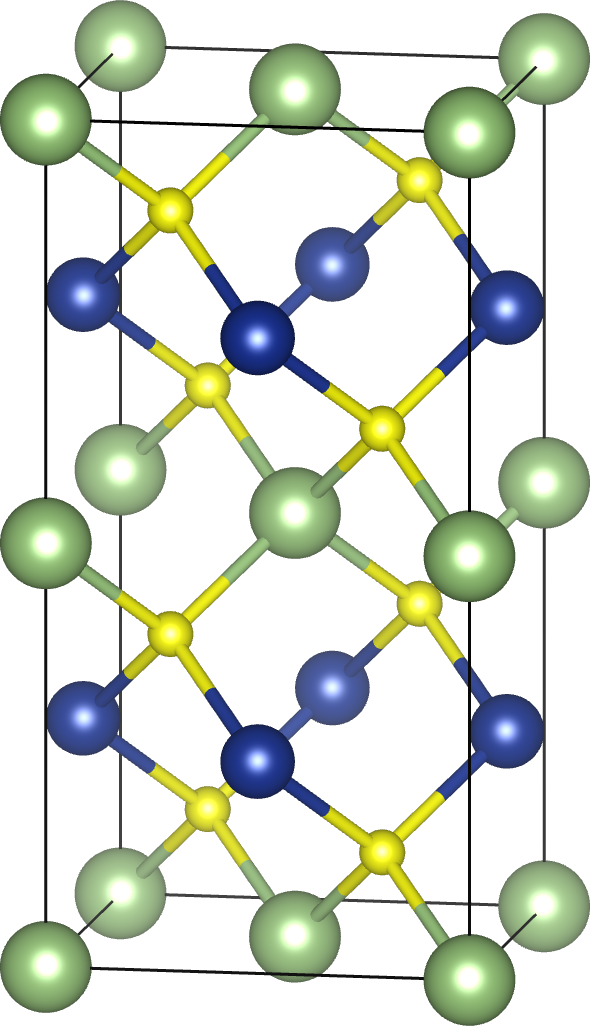}
\caption{}
\end{subfigure}
\vspace{0.7em}\\
\includegraphics[height=1.8em]{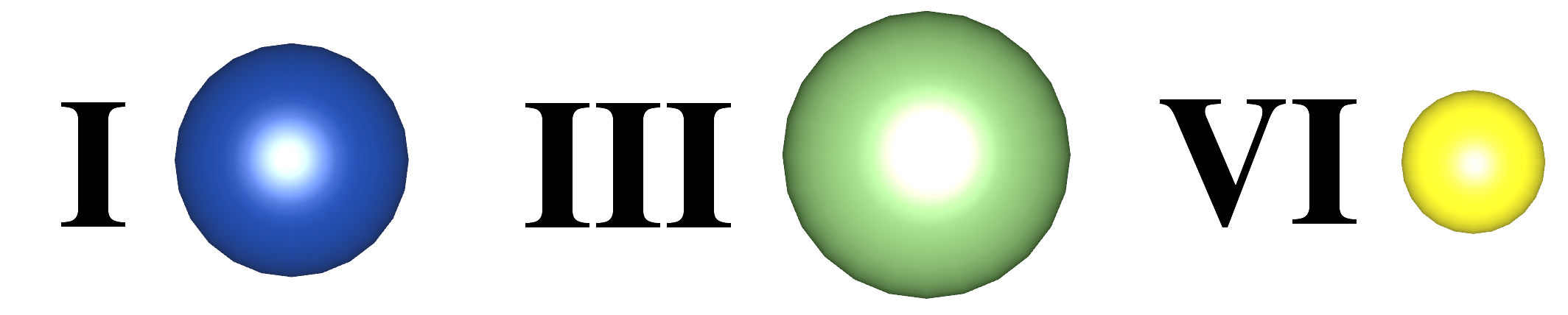}
\caption{\label{fig:structure} Chalcopyrite (a) and CuAu-like (b) structure of ternary \mbox{I-III-VI$_2$} compounds.}
\end{figure}

I-III-VI$_2$ compounds are stable at room temperature in the chalcopyrite (CH) structure (space group I$\bar{4}$2d). However, Su and Wei~\cite{Su1999} have used TEM to demonstrate the presence of CuAu-like (CA) orderings (space group P$\bar{4}$2m) in thin films of CuIn(S,Se)$_2$, grown by vapor-phase epitaxy on Si and GaAs substrates.  Alvarez et al.~\cite{Alvarez2002} also analyzed films of CuInS$_2$, using XRD to estimate the relative amount of phase domains. They found that the total amount of CA ordered phase in samples grown under Cu-poor conditions was between 8\% and 25\%. By growing films of CuInS$_2$ on various Si substrates, Su et al.~\cite{Su2000} discovered that although the CA phase is always present, the amount of CA domains is influenced by the substrate orientation. Moreover, Hahn et al.~\cite{Hahn2001} found that by using a Si(001) substrate, the CA phase will dominate the orderings of the cation sublattice. Recently, Moreau et al.~\cite{Moreau2015} have stated that for the CuInS$_2$ compound, introducing domains of CA phase can lead to a reduction of strain in the absorber layer, resulting in an increased carrier mobility and reduced recombination. Despite the fact that this phase is often found together with CH in thin films, little research has been done to determine its properties. Figure~\ref{fig:structure} shows the CH and CA structure of the ternary I-III-VI$_2$ materials.

In this paper we present a first-principles investigation of the efficiency of the CA phase for a selection of compounds. First, we analyze the thermodynamic stability in order to determine the likelihood of the presence of CA domains within a CH-based solar cell. We continue by presenting the optoelectronic properties of the CA phase materials. Next, we use these results to calculate the SLME and discuss the obtained efficiencies of specific compounds. For one of the compounds, the SLME is above the SQ limit. Similar results have been observed by Sarmadian et al.~\cite{Nasrin2016}. We analyze this surprising result in more detail by taking a closer look at how the calculated efficiency depends on the thickness and band gap of the material, as well as the temperature of the device. We deduce that the detailed balance approach for the radiative recombination current allows for higher open circuit voltages at lower thicknesses, producing a higher SLME than the corresponding SQ limit. In order to broaden our analysis to indirect band gap materials, we study the SLME of silicon. We find that in the SLME model, the fraction of non-radiative recombination is such that many indirect band gap absorber layers have a very high reverse saturation current, resulting in an unreasonably low calculated efficiency. 

\section{Computational Details}\label{sec:computational}

We make a selection of ten compounds for which we can compare the calculated efficiency of the CA phase with the CH results of Yu and Zunger~\cite{Yu2012}. The CA and CH structure are studied using a first-principles approach within the Density Functional Theory (DFT) formalism, as implemented in the Vienna Ab initio Simulation Package~\cite{Kresse1993,Kresse1996,Kresse1996b} (VASP). The Projector Augmented Wave (PAW) method~\cite{Blchl1994} is applied, and the electrons that are treated as valence electrons are underlined in Table~\ref{tab:valElec}. The exchange-correlation functional is calculated using the Generalized Gradient Approximation (GGA) of Perdew-Burke-Ernzerhof (PBE)~\cite{Perdew1996}. The energy cutoff for the plane wave basis is set to 350 eV, and a 4$\times$4$\times$4 Monkhorst-Pack~\cite{Monkhorst1976} (MP) mesh is used for sampling the first Brillouin zone. Electronic convergence is obtained when the energy difference between two electronic steps is smaller than $10^{-4}$~\si{\electronvolt}. The structure is considered converged when the forces on the atoms are all below $10^{-2}$~\si{\electronvolt}/\si{\angstrom}. 

\begin{table}[htbp]
\centering
\setlength{\captionmargin}{20pt}
\renewcommand{\arraystretch}{1.2}
\caption{\label{tab:valElec}Electron configuration of the atoms.}
\begin{tabular}{c@{\hskip 2 em}l}\hline
Element & Configuration \\\hline
Cu & [Ar] \underline{3d$^{10}$4s$^1$} \\
Ag &[Kr] \underline{4d$^{10}$5s$^1$} \\
Ga &[Ar] 3d$^{10}$\underline{4s$^2$4p$^1$}\\
In &[Kr] 4d$^{10}$\underline{5s$^{2}$4p$^1$} \\
S &[Ne] \underline{3s$^2$3p$^4$} \\
Se &[Ar]  3d$^{10}$\underline{4s$^{2}$4p$^4$} \\
Te &[Kr] 4d$^{10}$\underline{5s$^2$5p$^4$}\\
\hline
\end{tabular}
\end{table}

Because an accurate band gap is important for the correct evaluation of the efficiency, we perform single shot G$_0$W$_0$~\cite{Hybertsen1985} calculations on top of HSE06~\cite{Heyd2006}. However, in order to accurately update the quasiparticle energies within the G$_0$W$_0$ approximation, it is necessary to consider the semi-core electrons as valence electrons within the PAW framework~\cite{Fuchs2007}. Hence, we treat the 3$s$, 3$p$ and 3$d$ (4$s$, 4$p$ and 4$d$) orbitals as valence states for the Ga (In) atoms for the G$_0$W$_0$@HSE06 calculations of the band gap. In addition, we use a well converged 8$\times$8$\times$8 MP mesh, an increased energy cutoff of 400~\si{\electronvolt} and a large amount of unoccupied bands (600 in total). 

\begin{table*}[htbp]
\centering
\renewcommand{\arraystretch}{1.5}
\caption{\label{tab:lattice}Lattice parameters of the \mbox{CuAu-like} (CA) and chalcopyrite (CH) phase of the considered compounds}
\begin{tabular*}{\textwidth}{@{\extracolsep{\fill}}lccccccccc}\hline
\multirow{2}{*}{Material}    & & CA & & & CH & & & CH(Ref~\cite{Hahn1953}) & \\ \cmidrule(lr){2-4} \cmidrule(lr){5-7} \cmidrule(lr){8-10}
 			&  $a$ (\si{\angstrom}) & $c$ (\si{\angstrom}) & $c/a$ & $a$ (\si{\angstrom}) & $c$ (\si{\angstrom}) & $c/a$ & $a$ (\si{\angstrom}) & $c$ (\si{\angstrom}) & $c/a$ \\\hline
AgGaSe$_2$ 	& 5.702 & 12.663 & 2.221 & 6.045 & 11.267 & 1.864 & 5.973 & 10.88 & 1.823 \\
AgGaTe$_2$ 	& 6.220 & 13.060 & 2.100 & 6.403 & 12.327 & 1.925 & 6.283 & 11.94 & 1.897 \\
AgInS$_2$  	& 5.780 & 12.132 & 2.100 & 5.925 & 11.554 & 1.950 & 5.816 & 11.17 & 1.920 \\
AgInTe$_2$ 	& 6.511 & 13.224 & 2.031 & 6.570 & 13.000 & 1.979 & 6.406 & 12.56 & 1.962 \\
CuGaS$_2$  	& 5.341 & 10.861 & 2.033 & 5.384 & 10.669 & 1.982 & 5.349 & 10.47 & 1.958 \\
CuGaSe$_2$ 	& 5.662 & 11.436 & 2.020 & 5.683 & 11.277 & 1.984 & 5.607 & 10.99 & 1.960 \\
CuGaTe$_2$ 	& 6.109 & 12.170 & 1.992 & 6.091 & 12.160 & 1.996 & 5.994 & 11.91 & 1.987 \\
CuInS$_2$ 	& 5.636 & 11.129 & 1.975 & 5.598 & 11.274 & 2.014 & 5.517 & 11.06 & 2.005 \\
CuInSe$_2$ 	& 5.914 & 11.710 & 1.980 & 5.881 & 11.840 & 2.013 & 5.773 & 11.55 & 2.001 \\
CuInTe$_2$ 	& 6.323 & 12.590 & 1.991 & 6.313 & 12.681 & 2.009 & 6.167 & 12.34 & 2.000 \\ \hline
\end{tabular*}
\end{table*}

The optical properties are calculated within the Random Phase Approximation (RPA), using the long wavelength expression for the imaginary part of the dielectric tensor~\cite{Gajdo2006,Harl2007}. The real part of the dielectric tensor is determined using the Kramers-Kronig relation\footnote{The Kramers-Kronig relation is calculated by VASP using a complex shift (``CSHIFT''). After calculating the real part, however, VASP also recalculates the corresponding imaginary part. Since the complex shift introduces a broadening, this causes an earlier onset of the imaginary part, and consequently in the absorption coefficient. In order to prevent this, we commented out the line in the VASP code that recalculates the imaginary part.}. In order to get an accurate description of the energy levels, the exchange-correlation energy is calculated with the HSE06 functional, which has been reported~\cite{Wan2013} to produce optical properties close to those obtained from experiment for CuIn(S$_x$Se$_{x-1}$)$_2$. We found that it is enough to sample the Brillouin zone using a 12$\times$12$\times$12 MP mesh to obtain a converged dielectric tensor. The number of unoccupied bands is increased to at least three times the number of occupied bands. Because of the tetragonal symmetry of the CA structure, the resulting dielectric tensor is diagonal and has two independent components $\varepsilon_{xx}$ (=$\varepsilon_{yy}$) and $\varepsilon_{zz}$. Since we make no assumptions about the direction from which the photons enter the absorber layer, we average the diagonal components to derive the dielectric function $\varepsilon (E) = \varepsilon^{(1)} (E) + i \varepsilon^{(2)} (E) $ at energy $E$. Finally, in order to obtain a more accurate onset of the absorption spectrum, we shift the imaginary part of the dielectric function to the G$_0$W$_0$@HSE06 band gap, and recalculate the real part using the Kramers-Kronig relations. 

Once we have acquired the dielectric function, we can calculate the absorption coefficient
\begin{equation}
	\alpha(E) = \frac{4 \pi E }{hc}\hat{k}(E),
\end{equation}
with $h$ Planck's constant and $c$ the speed of light, from the extinction coefficient $\hat{k}(E)$:
\begin{equation}
	\hat{k}(E) = \sqrt{\frac{|\varepsilon(E)| - \varepsilon^{(1)}(E)}{2}}.
\end{equation}
This allows us to determine the absorptivity $a(E) = 1 - e^{-2\alpha(E)L}$ for an absorber layer of thickness $L$ with a reflecting back surface~\cite{Green1981}. 

The theoretical maximum solar cell efficiency is defined as
\begin{equation}\label{eq:efficiency}
\eta = \frac{P_m}{P_{in}},
\end{equation}
where $P_m$ is the maximum power density and $P_{in}$ is the total incident power density from the solar spectrum. For the SLME, the maximum power density is derived using the $J$-$V$ characteristic of the solar cell:
\begin{equation}\label{eq:power}
		P = J V =  \left( J_{sc} - J_0 \left(e^{\frac{eV}{kT}} - 1 \right) \right) V,
\end{equation}
with $J$ the total current density, $V$ the potential over the absorber layer, $k$ Boltzmann's constant, $T$ the temperature of the device and $e$ the elementary charge. The short circuit current density $J_{sc}$ and the reverse saturation current density $J_0$ are calculated from the absorptivity $a(E)$ of the material, as well as the AM1.5G solar spectrum $I_{sun}(E)$ and the black-body spectrum $I_{bb}(E,T)$:
\begin{equation}\label{eq:currents}
\begin{aligned}
J_{sc} &= e \int_0^\infty a(E) I_{sun} (E) dE,
\\ J_0 &= \frac{J_0^r}{f_r} = \frac{e\pi}{f}\int_0^\infty a(E)I_{bb}(E,T)dE,
\end{aligned}
\end{equation}
where $J_0^r$ is the radiative recombination current density. The fraction of radiative recombination $f_r$ is modeled using a Boltzmann factor:
\begin{equation}\label{eq:fraction}
	f_r = \exp\left({\dfrac{E_g^{da}-E_g}{kT}}\right),
\end{equation}
where $E_g$ and $E_g^{da}$ are respectively the fundamental and direct allowed band gap. 

\section{Structure and Formation energy}

To estimate the likelihood of finding a significant amount of CA domains in CuInSe$_2$, Wei et al.~\cite{Wei1999} used first-principles calculations to determine the difference in formation energy $\Delta E_f = E_{tot}^{CA} - E_{tot}^{CH}$ between the CH and CA phases of the compound. They found a very small energy difference of 2~\si{\milli\electronvolt}/atom, which led them to predict the coexistence of the CH and CA structures in CuInSe$_2$. This was confirmed experimentally by Su and Wei~\cite{Su1999}, supporting the idea that the presence of CA domains is a result of bulk thermodynamics. In order to determine the formation energy difference, we first optimize the structure of the CA and CH phase for each compound as described in Section~\ref{sec:computational}. 

\begin{table}[htbp]
\centering
\setlength{\captionmargin}{30 pt}
\renewcommand{\arraystretch}{1.2}
\caption{\label{tab:formation} Difference in formation energy between the chalcopyrite and CuAu-like structure of the considered ternary I-III-VI$_2$ compounds.}
\begin{tabular}{l@{\hskip 4 em}S[table-format=1.1]}
\hline
Material & {$\Delta E_f(\si{\milli\electronvolt}$/atom)} \\\hline
AgGaSe$_2$ & 31.3 \\
AgGaTe$_2$ & 27.8 \\
AgInS$_2$ & 8.9 \\
AgInTe$_2$ & 8.5 \\
CuGaS$_2$ & 8.8 \\
CuGaSe$_2$ & 9.9 \\
CuGaTe$_2$ & 7.0 \\
CuInS$_2$ & 1.6 \\
CuInSe$_2$ & 2.2 \\
CuInTe$_2$ & 2.9 \\ \hline
\end{tabular}
\end{table}

We show the calculated lattice parameters and $c/a$ ratio in Table~\ref{tab:lattice}, as well as the corresponding experimental values for the CH phase of the compounds\footnote[3]{No experimental values were found for the CA phase in the literature.}. We can see that the calculated $c/a$ ratios match well with those obtained from experiment. For the CA phase, replacing the cations Ag by Cu or Ga by In decreases the $c/a$ ratio of the unit cell. This trend is reversed for the CH phase. If we compare the $c/a$ ratio of the CA and CH phase, we find a large difference in the $c/a$ ratio for the \mbox{AgGa-VI$_2$} compounds. Table~\ref{tab:formation} presents the difference in formation energy for the selected list of compounds. Our first-principles results for CuInS$_{2}$, CuInSe$_{2}$ and CuGaSe$_{2}$ correspond well with those of Su et al.~\cite{Su2000}. Similar to the results for the $c/a$ ratio, the choice of cations has a large influence on the difference in formation energy. From Table~\ref{tab:formation}, we can see that substituting either In by Ga or Cu by Ag increases the difference in formation energy of the two phases. This means that if we consider the existence of the CA phase to be controlled by bulk thermodynamics, we expect CA domains to be common in the \mbox{CuIn-VI$_{2}$} compounds, and less likely in the \mbox{AgGa-VI$_2$} ones. 

\section{Absorber layer efficiency}

For all of the investigated compounds, we find a direct band gap at the $\Gamma$-point. Table~\ref{tab:Eg} presents a comparison between the G$_0$W$_0$@HSE06 band gaps calculated for the CA and CH structures\footnote[4]{We did not take all of the CH phase band gaps from Yu and Zunger~\cite{Yu2012}, because of inconsistencies between the tabulated and plotted values for some compounds in this paper.}. We can see that the G$_0$W$_0$ calculated band gaps for the CA phase are lower than those of the CH phase for all compounds besides CuInS$_2$. Furthermore, the difference is smaller for the \mbox{I-III-S$_2$} structures compared to the \mbox{I-III-(Se,Te)$_2$} compounds. Table~\ref{tab:Eg} also contains the experimental band gaps of the CH phase of the compounds$^\ddag$. We can see that although the G$_0$W$_0$@HSE06 band gaps correspond quite well to the experimental values for some compounds, there are clear discrepancies for others. This could be a result of the sensitivity of chalcogenide band gaps to the anion displacement $u$ (see the supplementary information$^\dag$ for more details). As an example of the dielectric function, we show the result for \mbox{CA-CuInS$_2$} in Fig.~\ref{fig:diel_CuInS2}. The results for the other compounds can be found in the supplementary information$^\dag$.

\begin{table*}[htbp]
\renewcommand{\arraystretch}{1.3}
\centering
\caption{\label{tab:Eg}Experimental and calculated band gaps of the \mbox{CuAu-like}(CA) and chalcopyrite (CH) phase of the considered compounds.}
\begin{tabular}{l@{\extracolsep{2em}}cccD{.}{.}{-1}c}
\hline
\multirow{2}{*}{Material} & \multicolumn{2}{c}{CA} & \multicolumn{3}{c}{CH}	 \\ \cmidrule(lr){2-3} \cmidrule(lr){4-6}
         & $E_g^{HSE}$(\si{\electronvolt}) & $E_g^{G_0W_0}$(\si{\electronvolt}) & $E_g^{HSE}$(\si{\electronvolt}) & \multicolumn{1}{c}{$E_g^{G_0W_0}$(\si{\electronvolt})} & \multicolumn{1}{c}{$E_g^{exp}$(\si{\electronvolt})}\\ \hline
AgGaSe$_2$ & 0.84 & 1.41 & - & 1.80^a & 1.83$^b$\\
AgGaTe$_2$ & 0.46 & 0.95 & 1.14 & 1.71 & 1.1-1.3$^b$\\
AgInS$_2$  & 1.20 & 1.69 & - & 1.74^a & 1.87$^b$\\
AgInTe$_2$ & 0.53 & 0.92 & - & 1.23^a & 0.96-1.04$^b$\\
CuGaS$_2$  & 1.77 & 1.94 & - & 1.99^a & 2.41$^c$\\
CuGaSe$_2$ & 0.96 & 1.19 & 1.29 & 1.46 & 1.64$^c$\\
CuGaTe$_2$ & 0.77 & 1.06 & - & 1.47^a & 1.23$^b$\\
CuInS$_2$  & 1.14 & 1.13 & 1.14 & 1.05 & 1.53$^c$\\
CuInSe$_2$ & 0.59 & 0.58 & 0.67 & 0.66 & 1.04$^c$\\
CuInTe$_2$ & 0.76 & 0.94 & - & 1.03^a & 0.96$^b$\\ \hline
\multicolumn{5}{l}{$^a$ Ref.~\citenum{Yu2012}, $^b$ Ref.~\citenum{Jaffe1984}, $^c$ Ref.~\citenum{Alonso2001}}
\end{tabular}
\end{table*}

\begin{figure}[htbp]
	\centering
		\includegraphics[width=\linewidth]{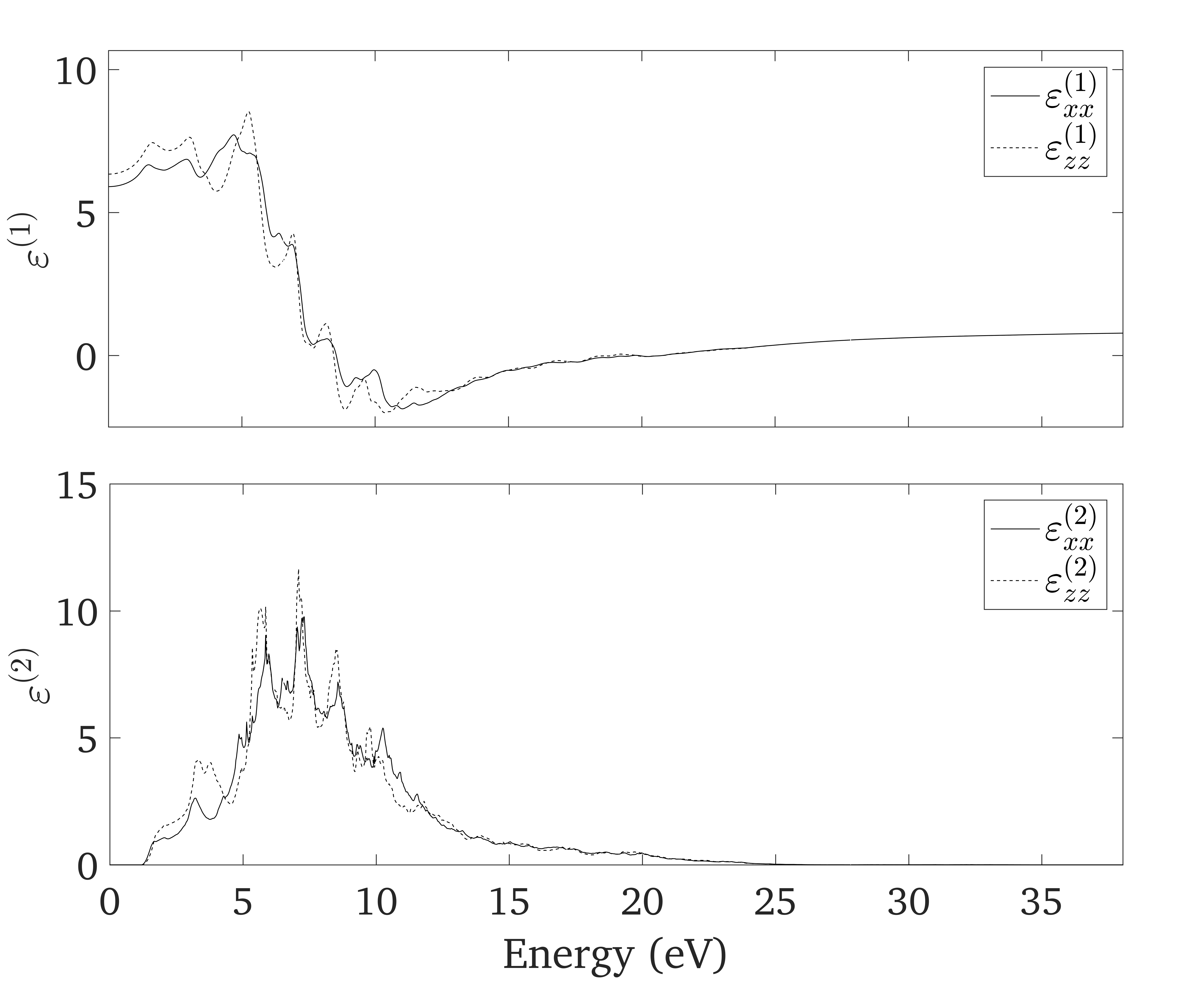}
	\caption{Real (upper figure) and imaginary (lower figure) parts of the calculated dielectric function of \mbox{CA-CuInS$_2$}.}
	\label{fig:diel_CuInS2}
\end{figure}

After we calculate the band gap and dielectric function for the CA phase of the selected list of compounds, we have all the required information to calculate their SLME. Because we find a direct allowed fundamental band gap for the CA phase of all of the compounds ($E_g^{da}=E_g$), we only have to consider cases where the non-radiative recombination is negligible ($f = 1$, see Eq.~(\ref{eq:fraction})). We present the calculated efficiency values in Table~\ref{tab:SLME}. In order to compare our results with those of Yu and Zunger, all efficiencies were calculated using thickness $L = 500$~\si{\nano\meter} and device temperature \mbox{$T = 300$~\si{\kelvin}}. 

\begin{table}[htbp]
\renewcommand{\arraystretch}{1.3}
\centering
\caption{\label{tab:SLME}Calculated SLME for both the \mbox{CuAu-like} and chalcopyrite~\cite{Yu2012} structures. The SQ limit of the corresponding band gap of the CA compounds is also given as a reference.}
\begin{tabular}{l@{\hskip 2em}c@{\hskip 1em}c@{\hskip 1em}c}
\hline
Material & SLME(\%) & SQ(\%) & SLME(\%)\\
		 &  (CA)	&  (CA) &  (CH)	 \\\hline
AgGaSe$_2$ & 27.0 & 33.3 & 15.8 \\
AgGaTe$_2$ & 28.9 & 31.1 & 21.8 \\
AgInS$_2$  & 23.1 & 29.1 & 19.7 \\
AgInTe$_2$ & 28.2 & 30.5 & 26.4 \\
CuGaS$_2$  & 16.4 & 24.1 & 16.5 \\
CuGaSe$_2$ & 27.8 & 33.4 & 26.6 \\
CuGaTe$_2$ & 28.9 & 32.0 & 24.8 \\
CuInS$_2$  & 29.0 & 33.5 & 23.1 \\
CuInSe$_2$ & 20.7 & 18.3 & 22.1 \\
CuInTe$_2$ & 27.9 & 30.9 & 28.0 \\ \hline
\end{tabular}
\end{table}

First, we note that several CH structures that are known to have high device efficiencies, such as CuIn(S,Se)$_{2}$, also have a high SLME. Moreover, it is clear that although the band gap has a large influence on the efficiency, some materials, such as CA- and \mbox{CH-AgInS$_2$}, have a very similar band gap but a significantly different calculated efficiency. This demonstrates the ability of the SLME to provide a more refined selection metric in comparison with the SQ limit. Finally, we see that for several compounds, the CA phase has a higher efficiency than the corresponding CH phase. This is consistent with the findings of Moreau et al.~\cite{Moreau2015}, who discovered that the presence of CA domains have a positive influence on the efficiency of CuInS$_2$. We suggest that the efficiency of these devices may have benefited from the presence of the CA phase directly through the optical properties of the material.

\begin{figure}[htbp]
	\setlength{\captionmargin}{5pt}
	\centering
		\includegraphics[width=\linewidth]{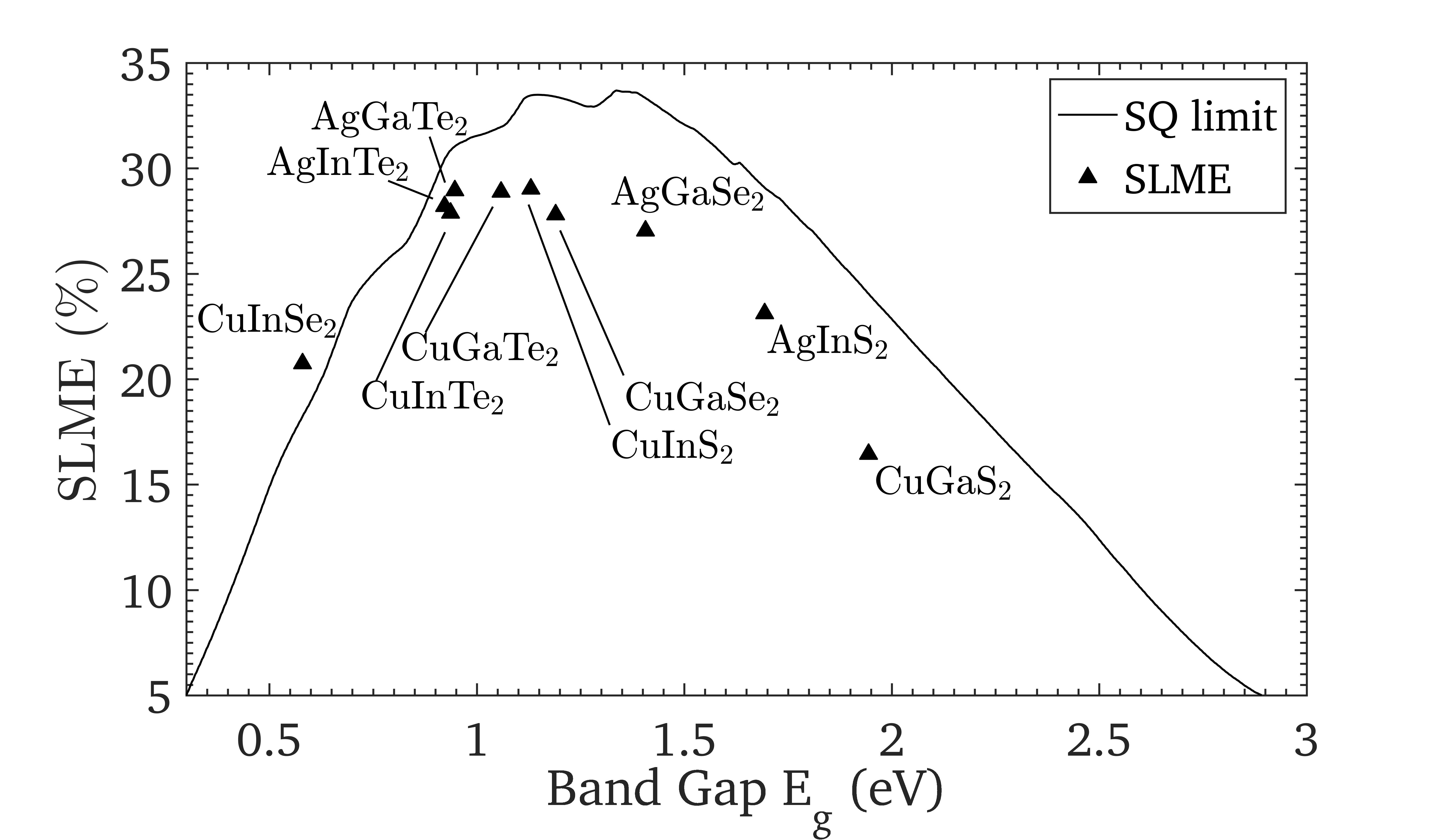}
	\caption{SLME of the CuAu-like compounds versus the band gap. All of the efficiencies were calculated using thickness $L = 500$ \si{\nano\meter} and device temperature $T = 300$~\si{\kelvin}. The black line represents the Shockley-Queisser limit.}
	\label{fig:SLME}
\end{figure}

\begin{figure}[htbp]
	\centering
		\includegraphics[width=\linewidth]{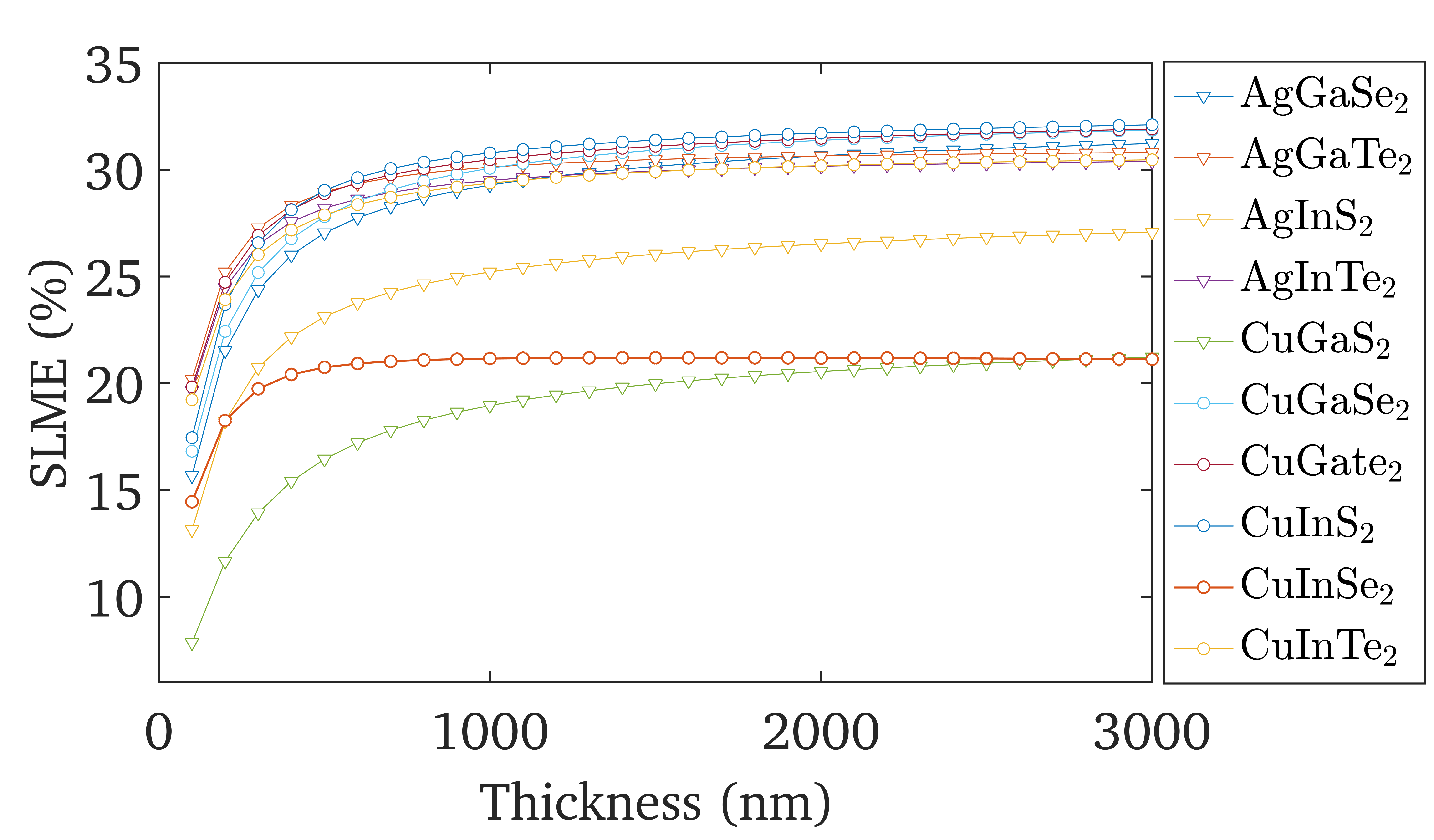}
	\caption{Calculated maximum efficiencies of the CuAu-like phase materials as a function of film thickness.}
	\label{fig:SLME_CuAu_L}
\end{figure}

In Fig.~\ref{fig:SLME}, we show the SLME of the CA phase of the various compounds versus their band gap, as well as the SQ limit. We immediately observe that the SLME value for \mbox{CA-CuInSe$_2$} is higher than the corresponding SQ limit. In Section~\ref{sec:Issues} we return to this result and discuss it in detail. The SLME is plotted as a function of the film thickness in Fig.~\ref{fig:SLME_CuAu_L}. We can see that for most compounds, the efficiency of the CA phase rises quickly for an increasing thickness. This demonstrates the potential of the CA phase compounds as absorber layers in thin-film solar cells. Finally, we discuss the issue of the possible discrepancy between the calculated and experimental band gaps for some of the compounds. Looking at Fig.~\ref{fig:SLME}, we expect the influence of the band gap to be small in the 1-1.5 \si{\electronvolt} interval. In case the calculated and experimental band gap are not in this region, however, any discrepancy between the calculated and experimental band gap is likely to influence the SLME.

\section{SLME Analysis\label{sec:Issues}}

\begin{figure}[htbp]
	\centering
		\includegraphics[width=\linewidth]{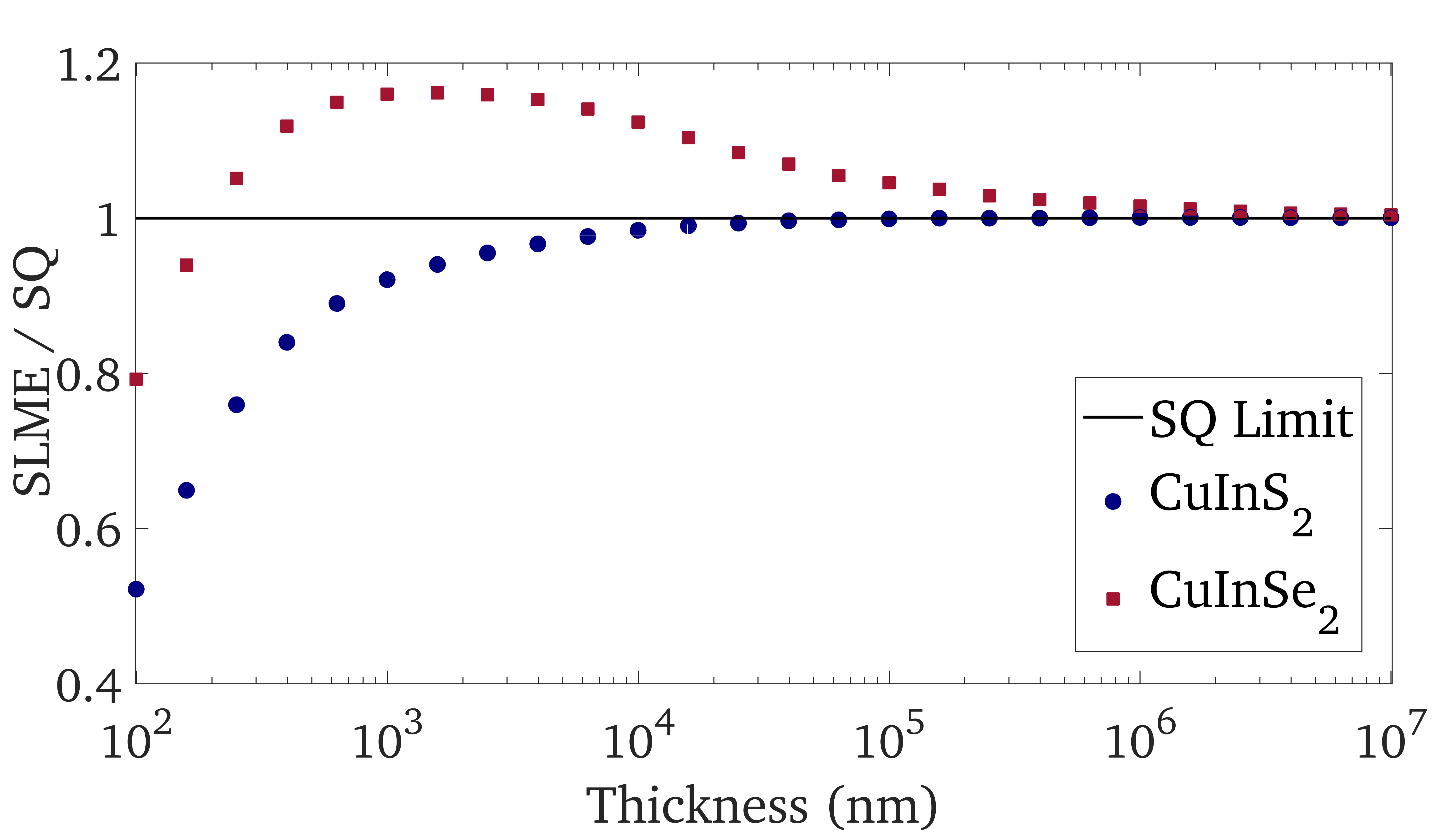}
	\caption{Thickness dependence of the SLME of CA-CuInSe$_2$ and CA-CuInS$_2$ at 300~\si{\kelvin} versus their SQ limit.}
	\label{fig:SLME_L}
\end{figure}

During the discussion of the SLME results of the CA phase, we noted that \mbox{CA-CuInSe$_2$} has an SLME value above the SQ limit. This result is surprising because the SQ limit is widely considered to be a theoretical maximum efficiency of a single junction absorber layer\footnote[5]{There are a number of considerations that could allow the efficiency of a single-junction solar cell to exceed the SQ limit, such as multiple-exciton generation, photon recycling, etc\ldots, however none of these are implemented by the SLME.}, and the SLME is based on the same detailed balance approach as the SQ limit. Due to the construction of the SLME, the calculated efficiency returns to the SQ limit for \mbox{$L \rightarrow \infty$}, since for an infinitely thick absorption layer the absorptivity becomes a step function. However, looking at the thickness dependence of the SLME for \mbox{CA-CuInSe$_2$} and \mbox{CA-CuInS$_2$} (Fig.~\ref{fig:SLME_L}), we see that the way they approach the SQ value is different. More specifically, the SLME of the compound CuInSe$_2$ crosses the SQ limit, whereas that of CuInS$_2$ does not.

\begin{figure}[htbp]
	\centering
		\includegraphics[width=\linewidth]{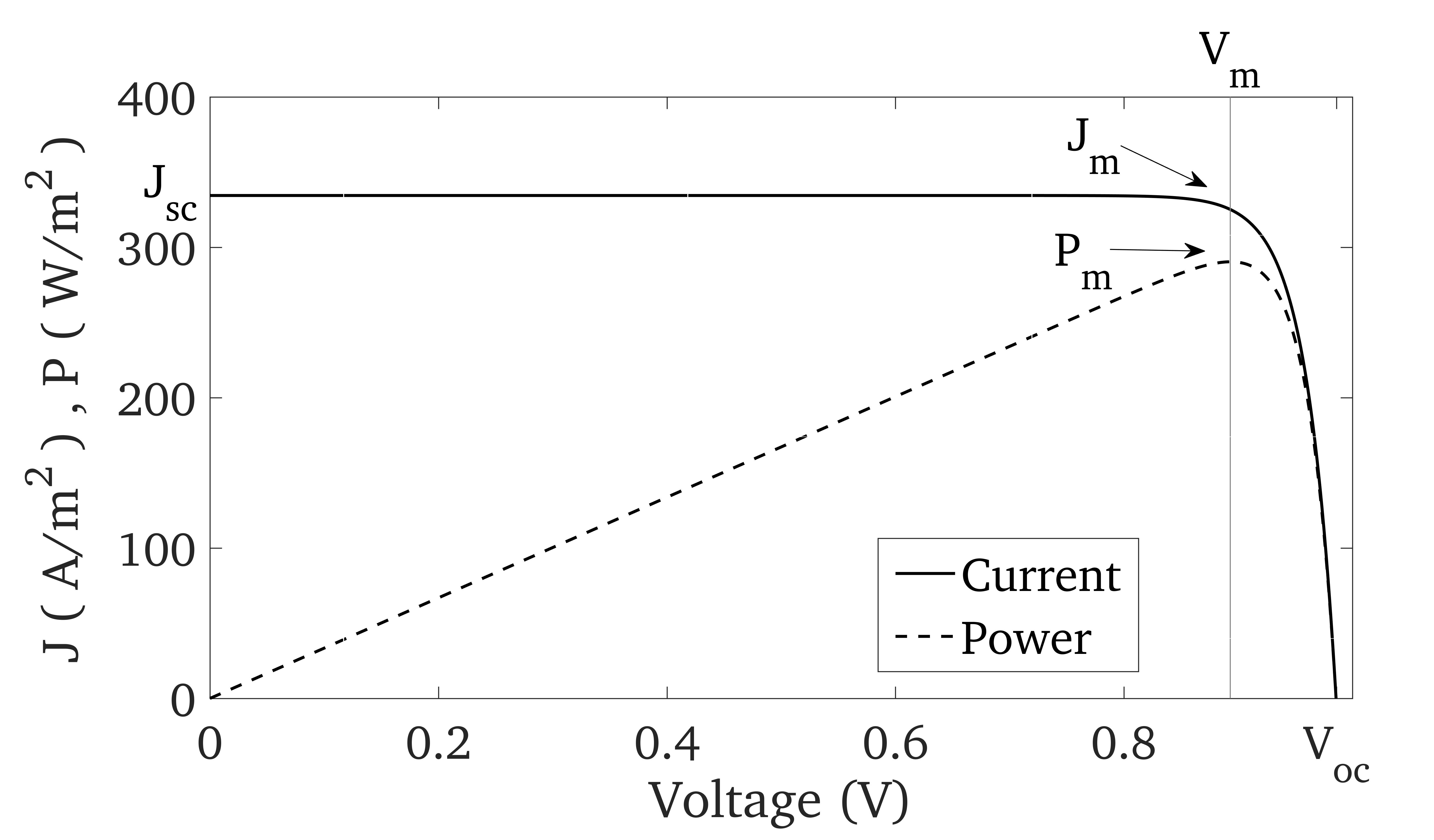}
	\caption{Calculated $J$-$V$ characteristic of CA-CuInS$_2$ at $T=300~\si{\kelvin}$ and $L = 500~\si{\nano\meter}$ (full line), as well as the corresponding power density (dashed line).}
	\label{fig:CuInS2_JV}
\end{figure}

\begin{figure}[!h]
	\centering
	\includegraphics[width=\linewidth]{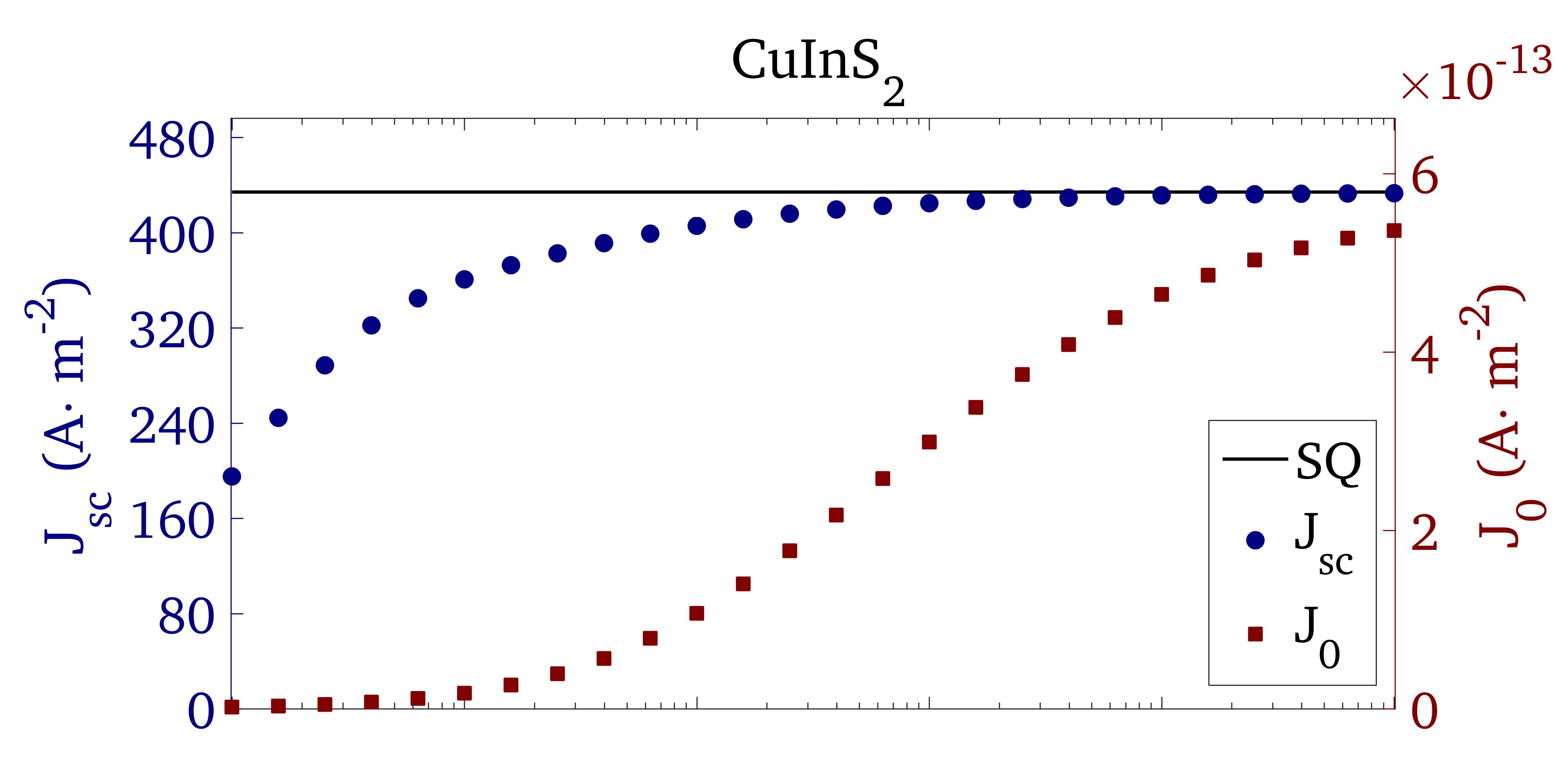}
	\includegraphics[width=\linewidth]{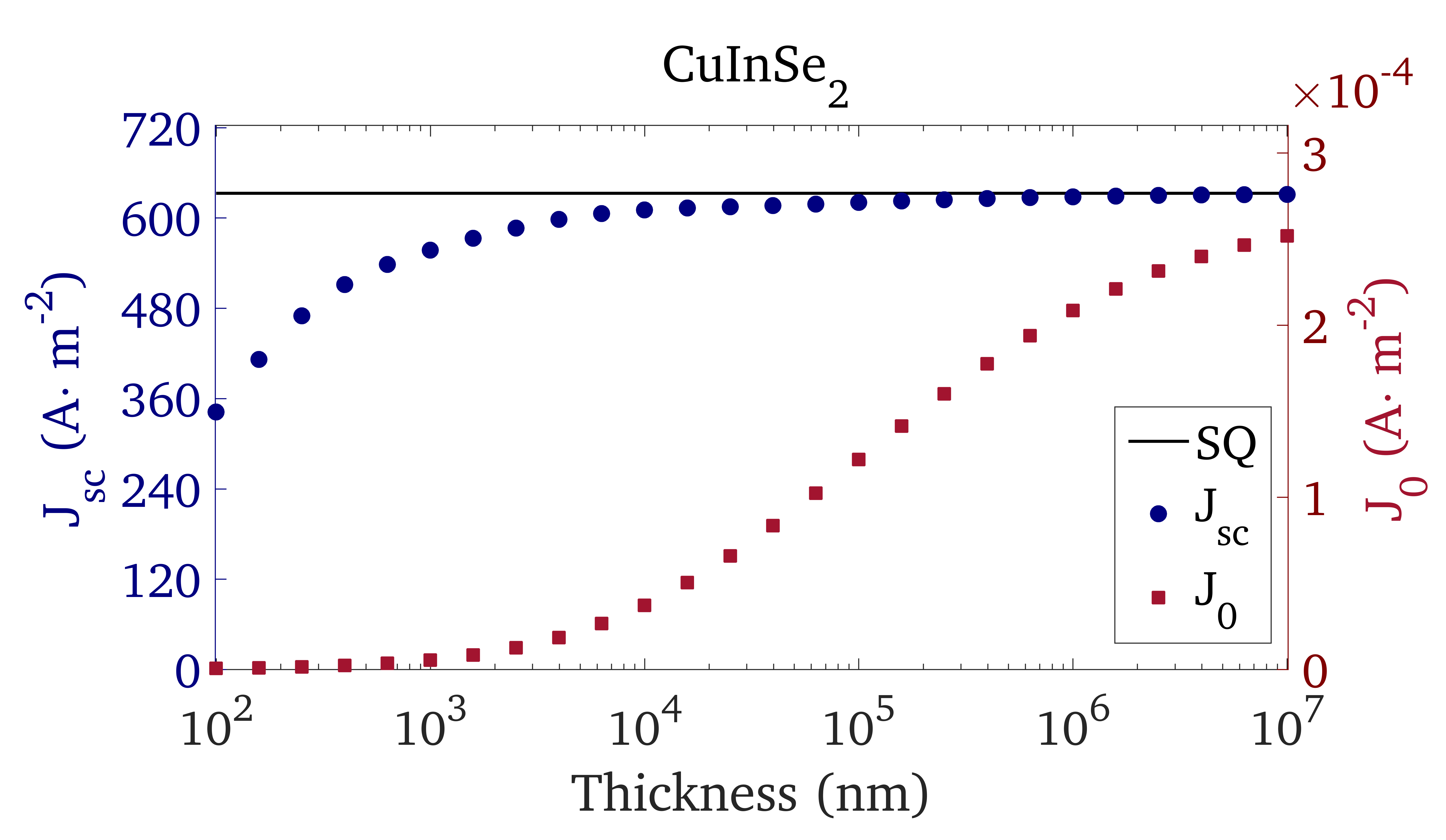}	
	\caption{Thickness dependence of the current densities of CuInS$_2$ (upper figure) and CuInSe$_2$ (lower figure) versus their respective SQ values.}
	\label{fig:J_L}
\end{figure}

We can understand the origin of this behavior by considering how the short-circuit current density $J_{sc}$ and reverse saturation current density $J_0$ are used to calculate the power density of the absorber layer (Eq.~(\ref{eq:power})). In Fig.~\ref{fig:CuInS2_JV} we show the calculated \mbox{$J$-$V$} characteristic of \mbox{CA-CuInS$_2$}. We can see that the total current density $J$ remains close to $J_{sc}$ up to a certain voltage. The value of this voltage, and hence the value of the open circuit voltage $V_{oc}$ and the voltage that maximizes the power density $V_{m}$, depends strongly on $J_0$. When we look at both current densities as a function of the thickness in Fig.~\ref{fig:J_L}, it is clear that for both compounds $J_{sc}$ converges to the corresponding SQ value much quicker than $J_0$. The relatively low value for $J_0$ at certain thicknesses allows for a higher open circuit voltage $V_{oc}$. This is the case for both \mbox{CA-CuInS$_2$} and \mbox{CA-CuInSe$_2$}. However, the order of magnitude of $J_0$ is much larger for CA-CuInSe$_2$ than for \mbox{CA-CuInS$_2$}.

\begin{figure}[htbp]
	\centering
		\includegraphics[width = 0.48\linewidth]{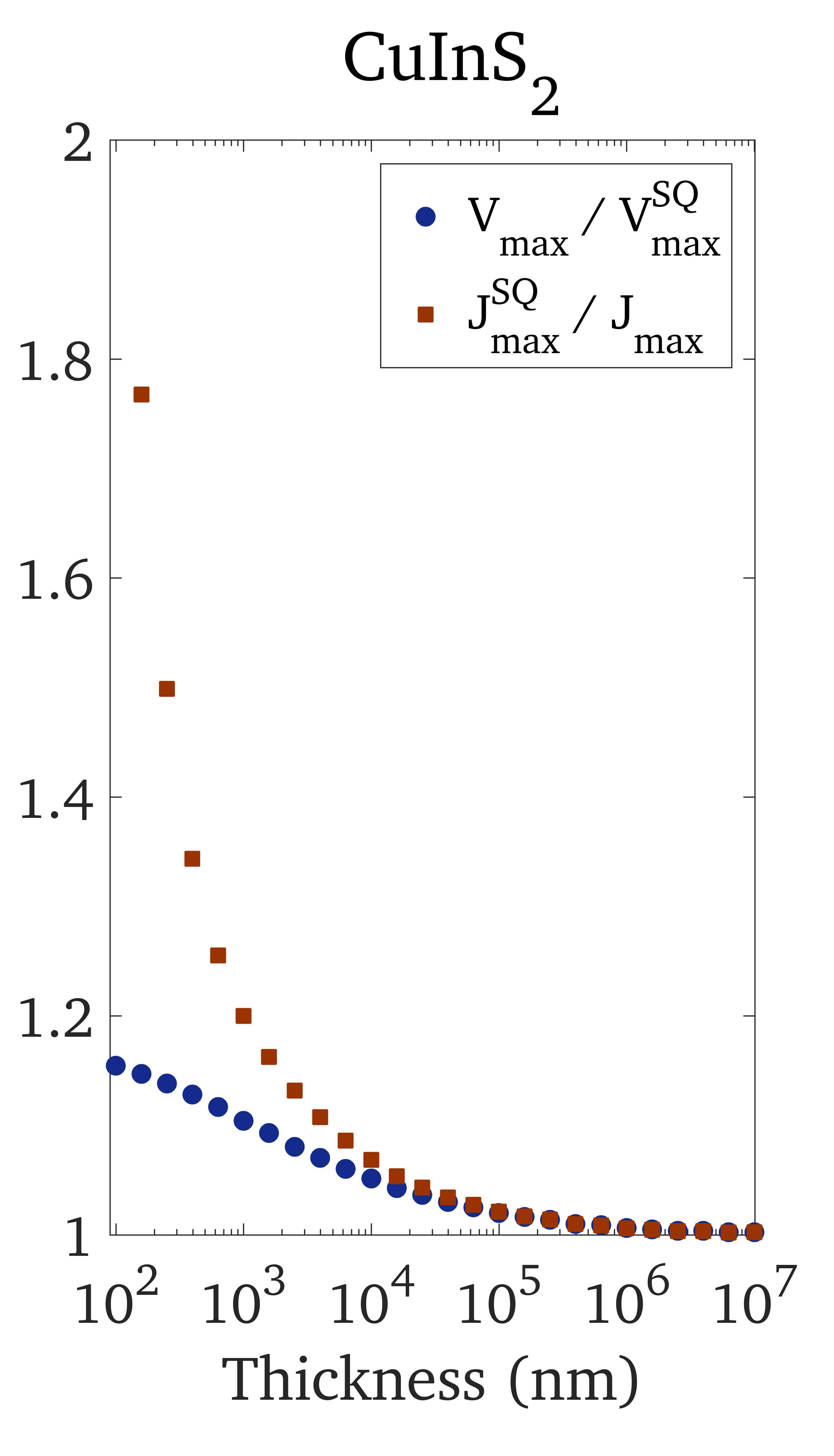}
		\includegraphics[width = 0.48\linewidth]{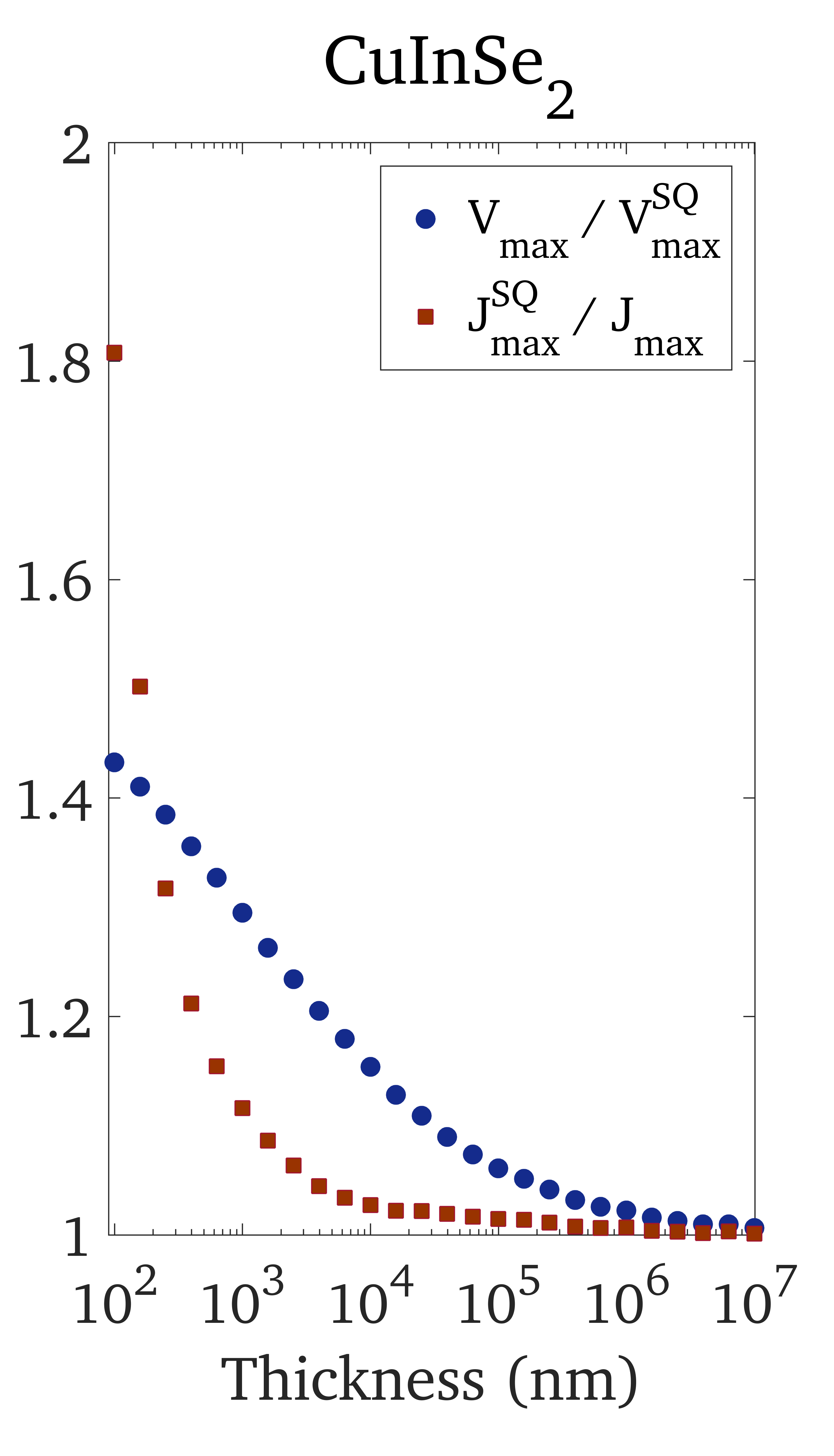}
	\caption{Comparison of the relative increase of the voltage that maximizes the power density ($V_{max}/V_{max}^{SQ}$) with the relative decrease of the corresponding current density ($J_{max}^{SQ}/J_{max}$).}
	\label{fig:VJcomp}
\end{figure}

The SLME crosses the SQ limit when its maximum power density is higher than the one calculated using the SQ values for $J_{sc}$ and $J_0$:
\begin{equation}\label{eq:aboveSQ}
\begin{aligned}
J_m V_m = P_m &> P_m^{SQ} = J_m^{SQ} V_m^{SQ} \\
\Leftrightarrow \hspace{0.4in}\frac{V_m}{V_m^{SQ}} &> \frac{J_m^{SQ}}{J_m}.
\end{aligned}
\end{equation}
Because the order of magnitude of $J_0$ is much larger for \mbox{CA-CuInSe$_2$}, the value of the fraction $V_m/V_m^{SQ}$ at low thicknesses is higher for \mbox{CA-CuInSe$_2$} when compared to that for \mbox{CA-CuInS$_2$} (Fig.~\ref{fig:VJcomp}). In comparison, the convergence of the fraction $J_m^{SQ}/J_m$ is similar for both compounds. From Eq.~(\ref{eq:aboveSQ}), it is clear that when $V_m/V_m^{SQ}$ is larger than $J_m^{SQ}/J_m$, the maximized power density is higher than its SQ value, which means that the SLME will be higher than the Shockley-Queisser limit for that thickness. Looking at Fig.~\ref{fig:VJcomp}, we can see that at $T=300$~\si{\kelvin}, this happens for CuInSe$_2$.

\begin{figure}[htbp]
	\centering
		\includegraphics[width=\linewidth]{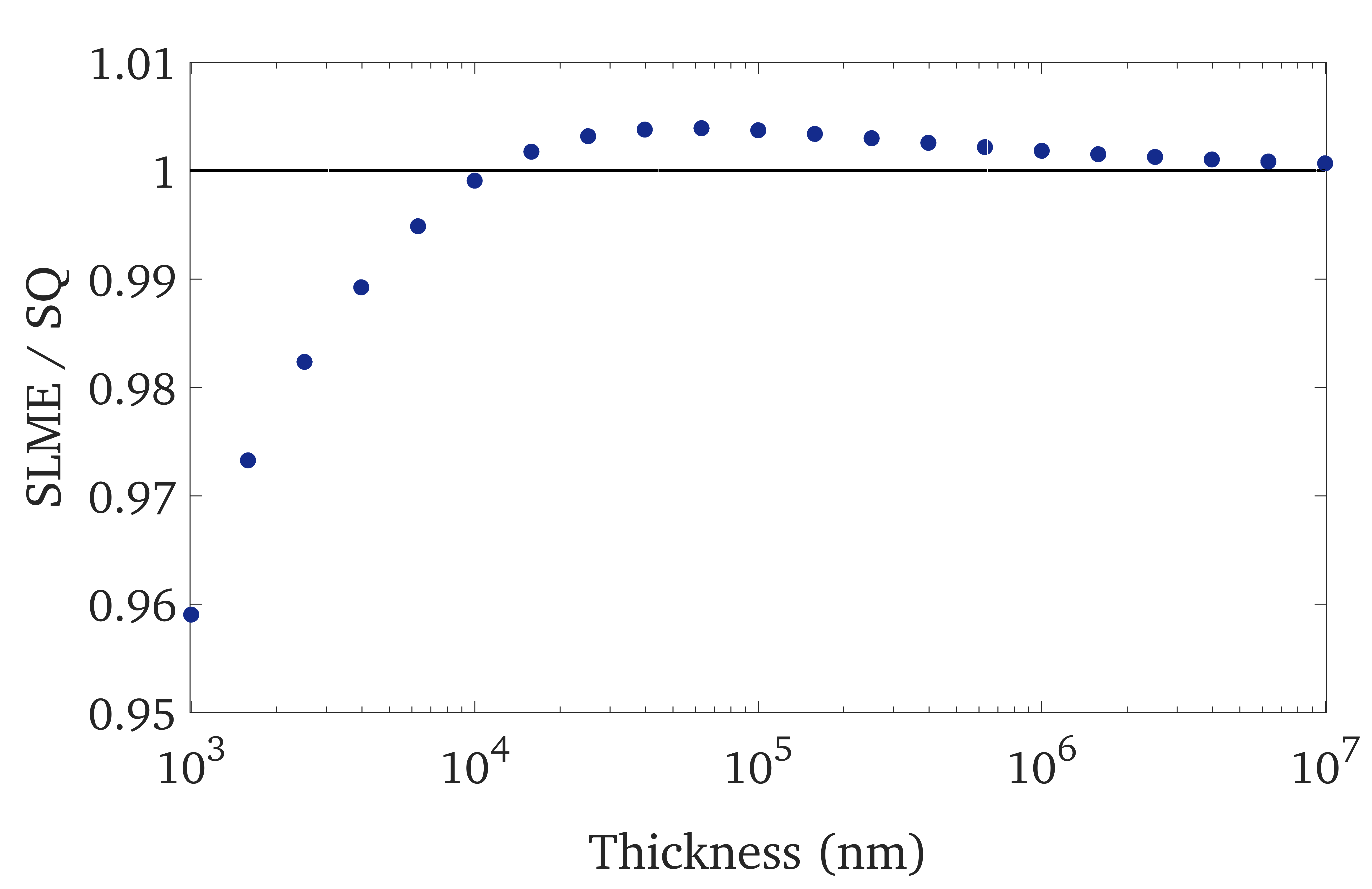}
	\caption{Thickness dependence of the SLME of CA-CuInS$_2$ at $T = 450$\si{\kelvin}.}
	\label{fig:SLME_highT}
\end{figure}

For direct band gap absorbers, $f_r = 1$, and $J_0$ is calculated from the overlap of the black-body spectrum $I_{bb}(E,T)$ and the absorptivity $a(E)$ of the material. From Eq.~(\ref{eq:currents}), we can understand that lowering the band gap increases $J_0$. As a result, materials with a low band gap are more likely to have an SLME value above the SQ limit at a specific thickness. It is also clear, however, that $J_0$ increases at higher temperatures. This raises the relative increase of $V_{max}$ at lower thicknesses, potentially producing calculated efficiencies above the SQ limit. For example, looking at the thickness dependence of the SLME of CA-CuInS$_2$ at $T=450\si{\kelvin}$ (Fig.~\ref{fig:SLME_highT}), we see that at this temperature the calculated efficiency also crosses the Shockley-Queisser limit.
 
Since the calculation of the SLME only deviates from the SQ limit by the introduction of an \textit{ab initio} calculated absorption spectrum, these results show that the SQ limit is not a theoretical upper limit within the assumptions of the detailed balance approach. This is because considering an infinite thickness for the solar cell, i.e. taking a step function for $a(E)$, overestimates $J_0$ as it is calculated in the detailed balance framework. As a result, it is possible that for a material with a certain band gap and absorptivity, $J_0$ is very low compared to its SQ value, which allows for a high $V_{oc}$. In case $V_{oc}$ is increased sufficiently, the total power density can go above that of the SQ limit, even though the calculated $J_{sc}$ is lower than its SQ value. In other words, if we consider all of the assumptions made in the Shockley-Queisser approach and introduce an absorption spectrum, it is possible to obtain efficiencies above the SQ limit.

\section{Indirect band gap absorbers}

\begin{figure}[htbp]
	\centering
		\includegraphics[width=\linewidth]{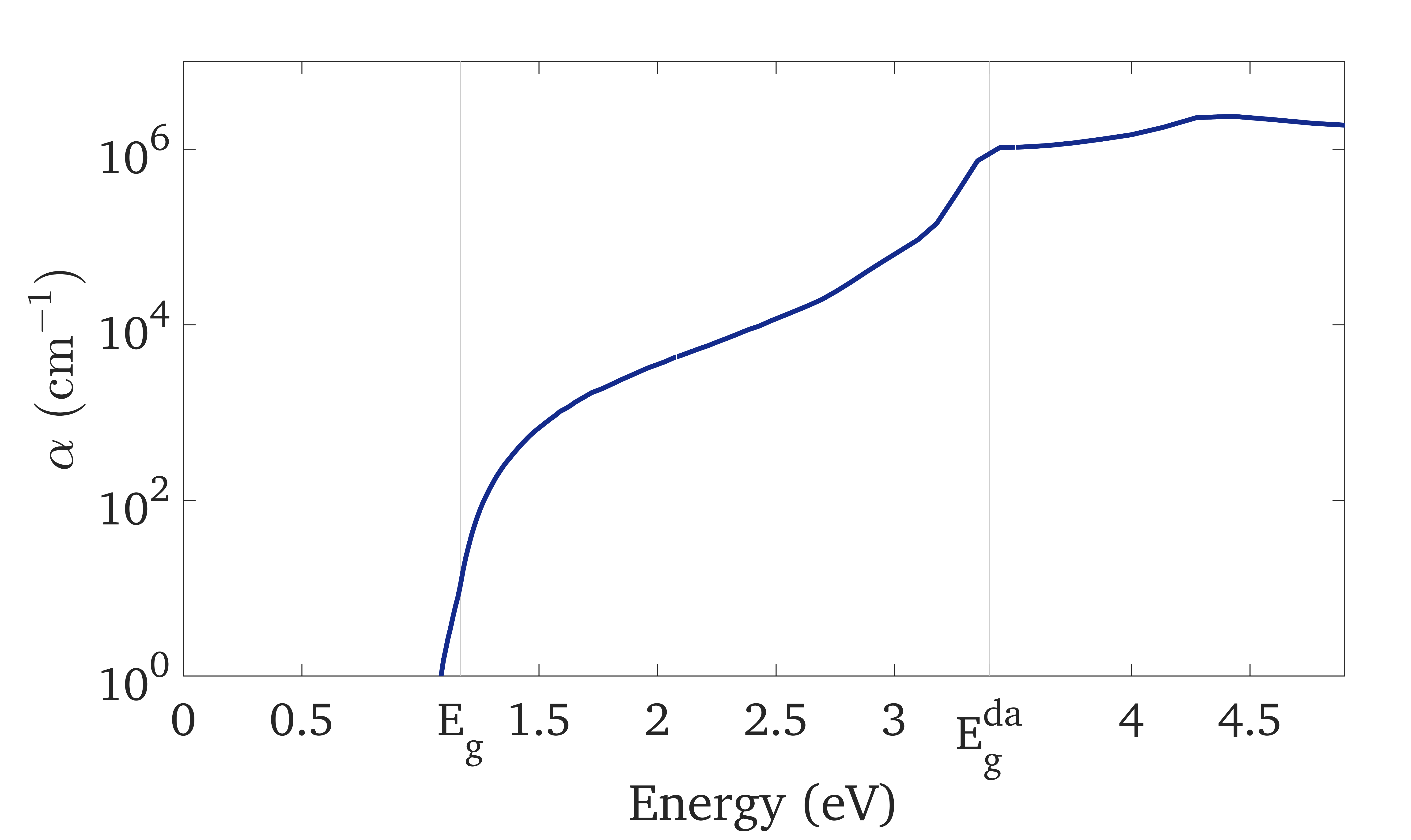}
	\caption{Experimental absorption coefficient at $T = 300\si{\kelvin}$ of crystalline silicon with data taken from~\cite{green2008}.}
	\label{fig:Si_expAbs}
\end{figure}

So far, we have only considered materials which have a direct band gap. For completeness, we expand our analysis to indirect band gap absorbers. We choose to calculate the SLME of silicon, which is currently the material that is still used the most for the production of solar cells. In Fig.~\ref{fig:Si_expAbs}, we show the experimental\footnote[6]{We choose to use an experimental spectrum in order to include the phonon-mediated contributions to the absorption coefficient.} absorption coefficient of crystalline silicon~\cite{green2008}. Notice the onset of the indirect and direct absorption at \mbox{$E_g = 1.17~\si{\electronvolt}$} and \mbox{$E_g^{da} = 3.4~\si{\electronvolt}$}, respectively.

\begin{figure}[htbp]
	\centering
		\includegraphics[width=\linewidth]{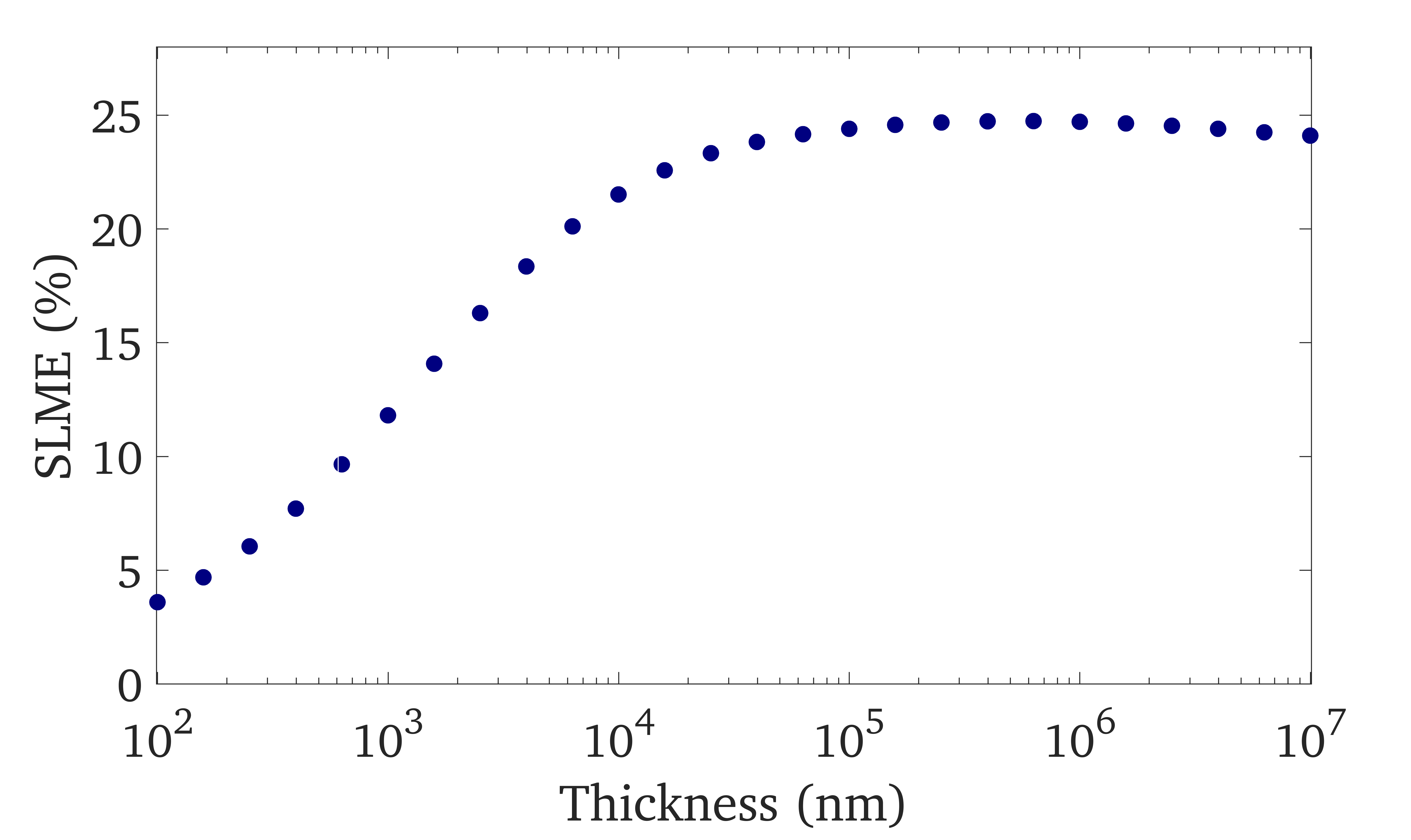}
	\caption{Thickness dependence of the SLME of silicon.}
	\label{fig:SLME_Si}
\end{figure}

Calculating the SLME using this optical spectrum produces an efficiency of zero for any value of $L$ and $T$. The origin of this troubling result is rooted in the fraction of radiative recombination expressed in Eq.~(\ref{eq:fraction}). Because of the large difference between the direct allowed and fundamental band gap of silicon ($\Delta = E_g^{da}-E_g=2.23$~\si{\electronvolt}), the radiative fraction is of the order $10^{-38}$. Since this fraction is used to calculate the reverse saturation current (see Eq.~(\ref{eq:currents})), this results in a $J_0$ that is unreasonably large. As discussed in Section~\ref{sec:Issues}, $J_0$ has a significant influence on the open circuit voltage $V_{oc}$. In this case, the high value of $J_0$ leads to a $V_{oc}$ that is too small to produce any significant power density. However, in case we set \mbox{$f_r = 10^{-3}$}, a more reasonable value for silicon~\cite{Shockley1952,Trupke2003,Richter2012}, then we obtain the results shown in Fig.~\ref{fig:SLME_Si}.

One could argue that silicon is a special case, and that generally efficient indirect absorbers do not have such a large band gap difference \mbox{$\Delta = E_g - E_g^{da}$}. For thin-film solar cells, indirect absorption also contributes significantly less to the power density. Consequently, indirect band gap materials with a large fundamental band gap are not suitable for these applications in any case. However, even for materials with a small $\Delta$, the modeled fraction of radiative recombination quickly becomes minute. For example, consider the compound Cu$_3$TlSe$_2$, which has been investigated by Yu and Zunger~\cite{Yu2012}. The reported difference between the fundamental and direct allowed band gap is $0.24$~\si{\electronvolt}. At 300~\si{\kelvin}, the fraction of radiative recombination then becomes \mbox{$f_r = 10^{-4}$}. This means that although $99.99$~\% of the recombination is non-radiative in nature, the reverse saturation current is still derived from an entirely radiative principle, based on the black-body spectrum in Eq.~(\ref{eq:currents}). Furthermore, it is clear that because of the exponential function in Eq.~(\ref{eq:fraction}), the fraction of radiative recombination drops very rapidly with increasing $\Delta$. This indicates that even for materials with a relatively low $\Delta$, the reverse saturation current will rise significantly, which is detrimental for the calculated efficiency. Hence, it is fair to question whether the recombination model of the SLME metric does not judge indirect band gap absorbers unfairly, potentially eliminating good materials during the selection procedure.\\

\section{Summary and Conclusions}

We have compared the structural and thermodynamic properties of the CA and CH phase of the compounds. By analyzing the difference in formation energy of the CH and CA phase, we conclude that CA domains are most likely to be present in \mbox{CuIn-VI$_2$} compounds, which is in good agreement with experimental results. From the calculated optoelectronic properties of the materials, we have determined their potential as absorbers for solar cells by applying the SLME selection metric. We identify several compounds with a high theoretical efficiency in the CA phase, most notably \mbox{CA-CuInS$_2$}, which has a significantly higher efficiency than the corresponding CH phase.

After observing an SLME value above the Shockley-Queisser limit for \mbox{CA-CuInSe$_2$}, we have performed a detailed analysis to find the origin of this result. We find that, within the details balance approach, the reverse saturation current $J_0$ approaches its SQ value very slowly for an increasing thickness $L$. This causes the SLME to cross the SQ limit for materials with a $J_0$ that is relatively high, i.e. materials with a low band gap or at higher temperatures. In other words, because the SQ limit overestimates $J_0$, it is not an effective theoretical maximum efficiency of a single junction cell within the detailed balance approach. Finally, we show that the model that introduces non-radiative recombination to the SLME quickly undercuts the efficiency of indirect band gap absorbers. 

\section{Acknowledgements}

We acknowledge financial support of FWO-Vlaanderen through projects G.0150.13N and G.0216.14N and ERA-NET RUS Plus/FWO, Grant G0D6515N. The computational resources and services used in this work were provided by the
VSC (Flemish Supercomputer Center) and the HPC infrastructure of the University of Antwerp (CalcUA), both
funded by the FWO-Vlaanderen  and the Flemish Government-department EWI.



\balance


\bibliography{ArXiv.bib} 
\bibliographystyle{rsc} 

\pagebreak
\onecolumn
\section*{Supplementary Information}

For some compounds, the calculated band gap of the chalcopyrite (CH) phase does not correspond well to the experimental value. One possible reason for these discrepancies is the sensitivity of the band gap of chalcogenides to the anion displacement $u$. Jaffe and Zunger~\cite{Jaffe1984} used standard DFT to demonstrate the influence of $u$ on the calculated band gap, finding that an increased $u$ leads to higher band gaps for CH-CuInSe$_2$ and CH-CuAlS$_2$. Similar results were found by Vidal et al.~\cite{Vidal2010} using HSE06 to calculate the band gap of CH-CuInS$_2$. In Table~\ref{tab:uPara}, we present the $u$ parameter calculated using PBE versus a set of experimental results. We can see that although $u$ corresponds well to experiment for some compounds, there are significant differences for others. For example, the calculated $u$ of both CH-CuInS$_2$ and CH-CuInSe$_2$ are below the experimental range. Considering the influence of $u$ on the band gap, it is not unreasonable to assert that the underestimation of the band gap may be related to the low values found for $u$.

\begin{table}[htbp]
\centering
\caption{\label{tab:uPara} Calculated and experimental~\cite{Jaffe1984} anion displacement for the chalcopyrite phase of the studied compounds.}
\begin{tabular}{l@{\hskip 2em}D{.}{.}{-1}@{\hskip 1em}D{.}{.}{4}}
\hline
Material   & \multicolumn{1}{c}{$u_{PBE}$} & \multicolumn{1}{c}{$u_{exp}$} \\\hline
AgGaSe$_2$ & 0.278 & 0.27   \\
		   &       & 0.276  \\\\
AgGaTe$_2$ & 0.266 & 0.26   \\\\
AgInS$_2$  & 0.256 & 0.25   \\
		   &       & 0.250  \\\\
AgInTe$_2$ & 0.243 & 0.25 \\\\
CuGaS$_2$  & 0.237 & 0.25   \\
		   &       & 0.275  \\
		   &       & 0.2539 \\
		   &       & 0.272  \\\\
CuGaSe$_2$ & 0.244 & 0.25   \\
		   &       & 0.250  \\
		   &       & 0.243  \\\\
CuGaTe$_2$ & 0.237 & 0.25   \\\\
CuInS$_2$  & 0.218 & 0.20   \\
           &       & 0.214  \\
		   &       & 0.2295 \\\\
CuInSe$_2$ & 0.217 & 0.22   \\
		   &       & 0.224  \\
		   &       & 0.235  \\\\
CuInTe$_2$ & 0.214 & 0.225  \\ \hline
\end{tabular}
\end{table}

For the calculation of the SLME, we require the dielectric tensor of the studied material. For all of the structures considered in this paper, the dielectric tensor is found to be diagonal and has two independent components:
\begin{equation}
\varepsilon_{\alpha\beta} (E) = \left(
\begin{matrix}
\varepsilon_{xx}(E) & 0 & 0 \\
0 & \varepsilon_{xx}(E) & 0 \\
0 & 0 & \varepsilon_{zz}(E) \\
\end{matrix}
\right),
\end{equation}
where each component is imaginary (e.g. $\varepsilon_{xx}(E) = \varepsilon_{xx}^{(1)}(E) + i \varepsilon_{xx}^{(2)}(E)$). Figures~\ref{fig:Agdiel} and~\ref{fig:Cudiel} show both the real and imaginary part of the calculated tensor components for the CuAu-like phase of the studied compounds. For completeness, we show the optical dielectric constants in Table~\ref{tab:dielCte}. It is interesting to note that the choice of the anion (S,Se,Te) in the CuAu-like phase of the I-III-VI$_2$ compounds has a strong influence on the value of their optical dielectric constants. The computational details can be found in Section 2 of the main text.

\begin{table}[htbp]
\centering
\setlength{\captionmargin}{120 pt}
\caption{\label{tab:dielCte} Optical dielectric constants of the CuAu-phase of the studied compounds.}
\begin{tabular}{l@{\hskip 2 em}S[table-format=1.1]S[table-format=1.1]}
\hline
Material & {$\varepsilon_{xx}(\infty)$} & {$\varepsilon_{zz}(\infty)$} \\\hline
AgGaSe$_2$ & 6.37 & 6.66 \\
AgGaTe$_2$ & 8.71 & 8.61 \\
AgInS$_2$ & 5.24 & 5.53 \\
AgInTe$_2$ & 7.96 & 8.20 \\
CuGaS$_2$ & 5.72 & 6.09 \\
CuGaSe$_2$ & 6.99 & 7.34 \\
CuGaTe$_2$ & 9.00 & 9.40 \\
CuInS$_2$ & 5.90 & 6.34 \\
CuInSe$_2$ & 7.18 & 7.69 \\
CuInTe$_2$ & 8.42 & 8.99 \\ \hline
\end{tabular}
\end{table}

\begin{figure}[htbp] 
\captionsetup[subfigure]{labelformat=empty}
\setlength{\captionmargin}{20pt}
\centering
\begin{subfigure}{0.4\textwidth}
\centering
\hspace{3em}AgGaSe$_2$
\includegraphics[width=0.8\linewidth]{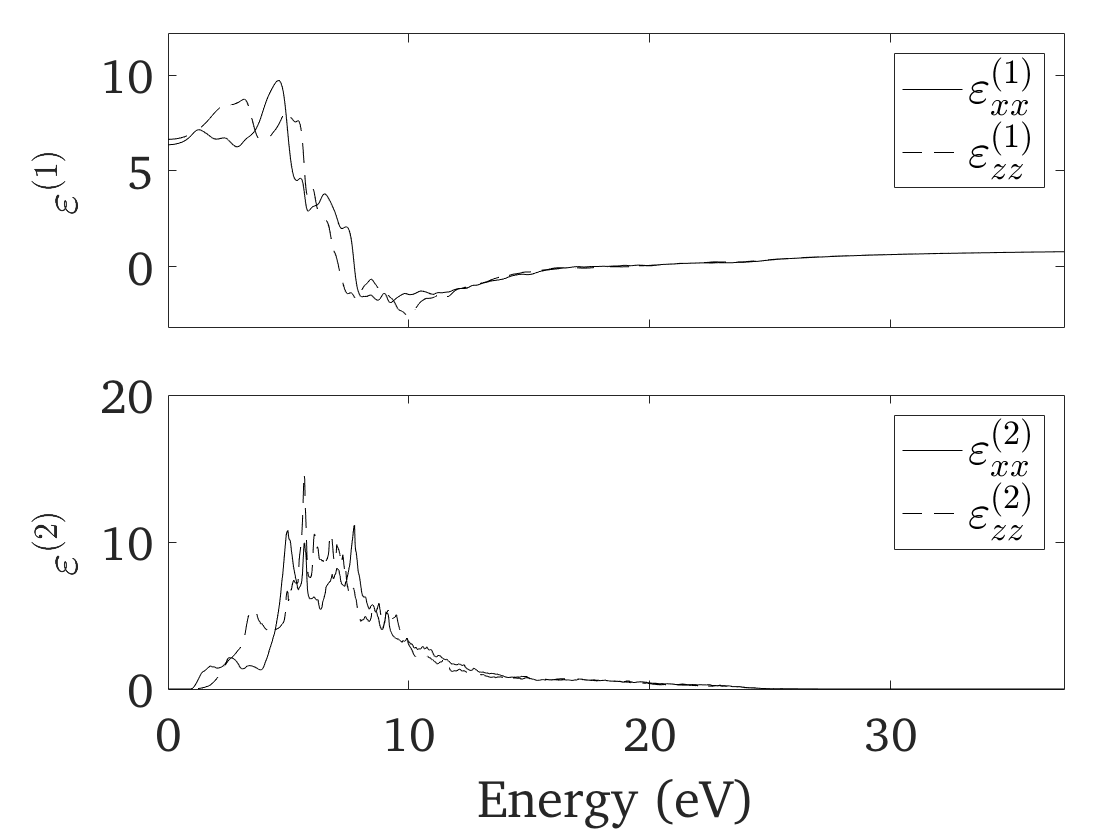}
\caption{}
\end{subfigure}%
\begin{subfigure}{0.4\textwidth}
\centering
\hspace{2em}AgGaTe$_2$
\includegraphics[width=0.8\linewidth]{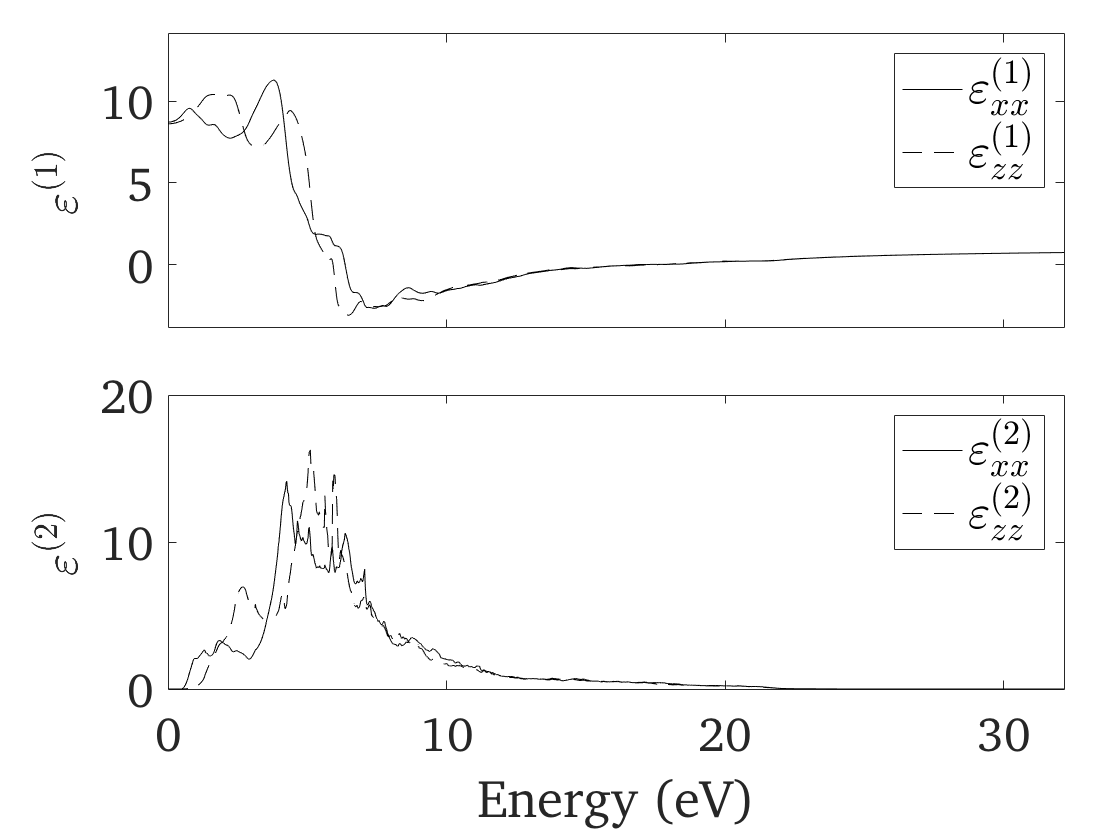}
\caption{}
\end{subfigure}
\begin{subfigure}{0.4\textwidth}
\centering
\hspace{3em}AgInS$_2$
\includegraphics[width=0.8\linewidth]{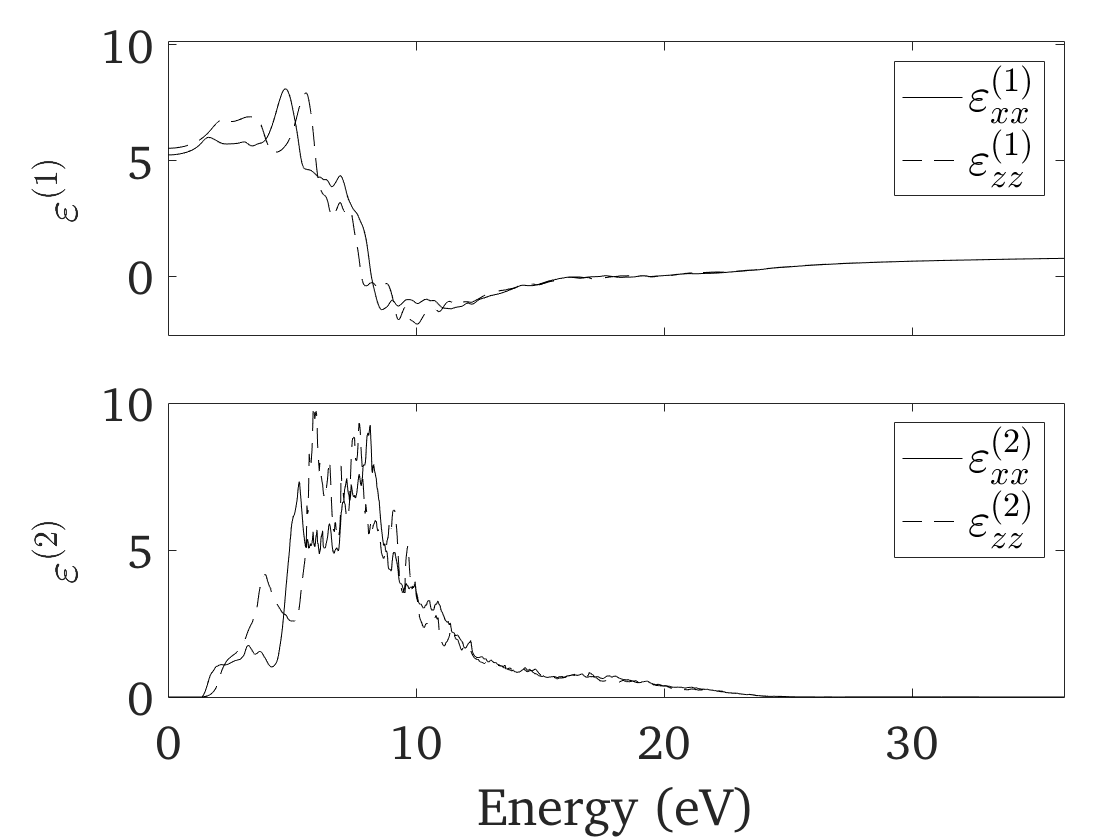}
\caption{}
\end{subfigure}%
\begin{subfigure}{0.4\textwidth}
\centering
\hspace{2em}AgInTe$_2$
\includegraphics[width=0.8\linewidth]{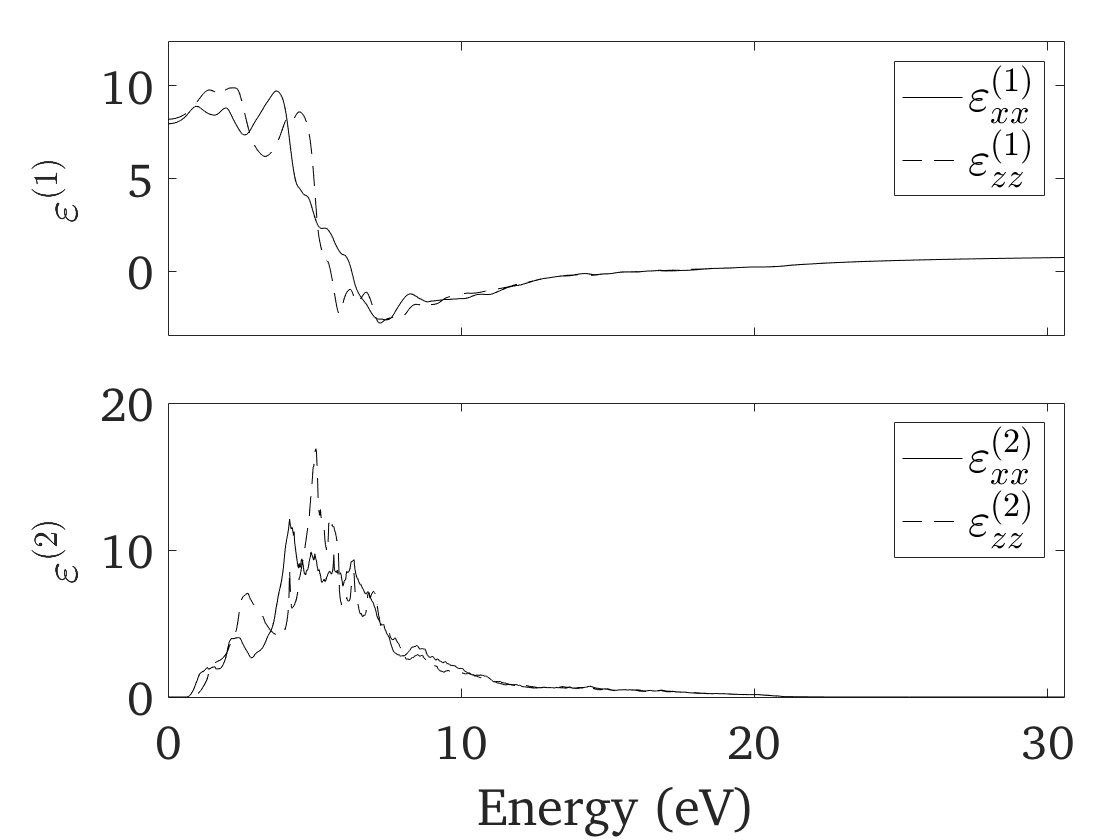}
\caption{}
\end{subfigure}
\caption{\label{fig:Agdiel}Dielectric functions of the CuAu-like phase of the Ag-III-VI$_2$ compounds.}
\end{figure}
	
\begin{figure}[htbp] 
\captionsetup[subfigure]{labelformat=empty}
\setlength{\captionmargin}{20pt}
\centering
\begin{subfigure}{0.4\textwidth}
\centering
\hspace{3em}CuGaS$_2$
\includegraphics[width=0.8\linewidth]{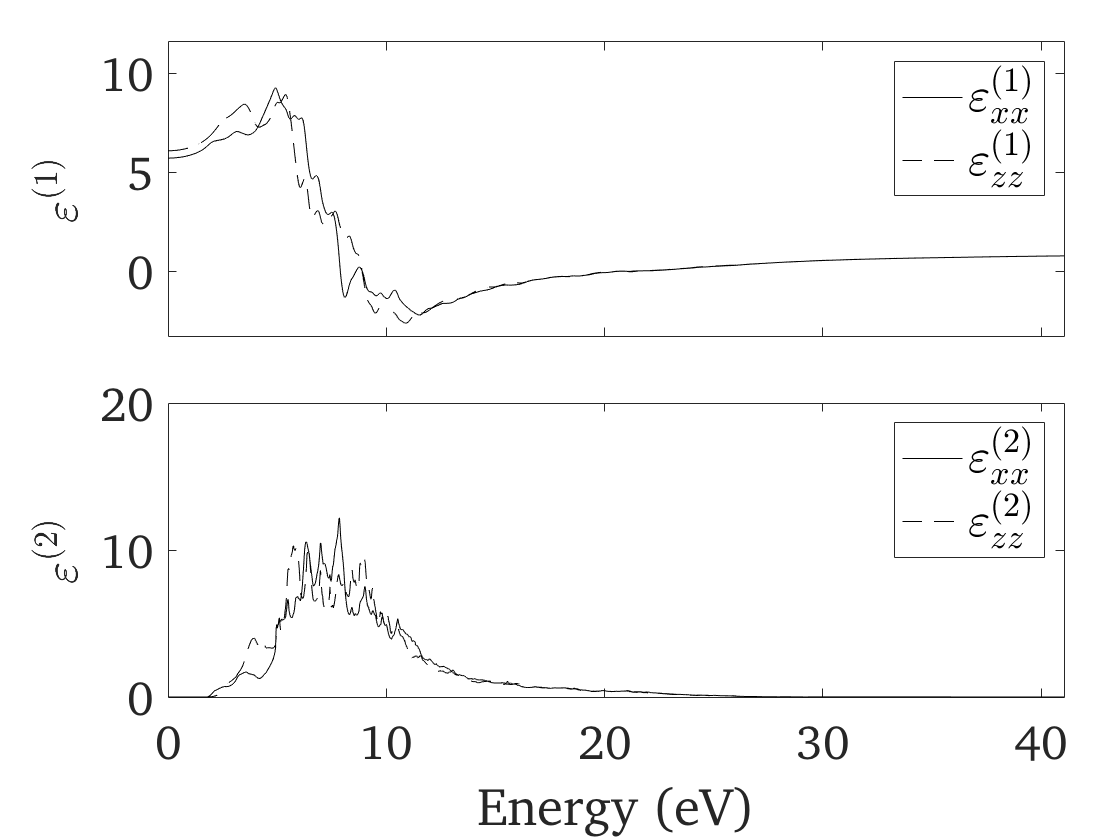}
\caption{}
\end{subfigure}%
\begin{subfigure}{0.4\textwidth}
\centering
\hspace{2em}CuGaSe$_2$
\includegraphics[width=0.8\linewidth]{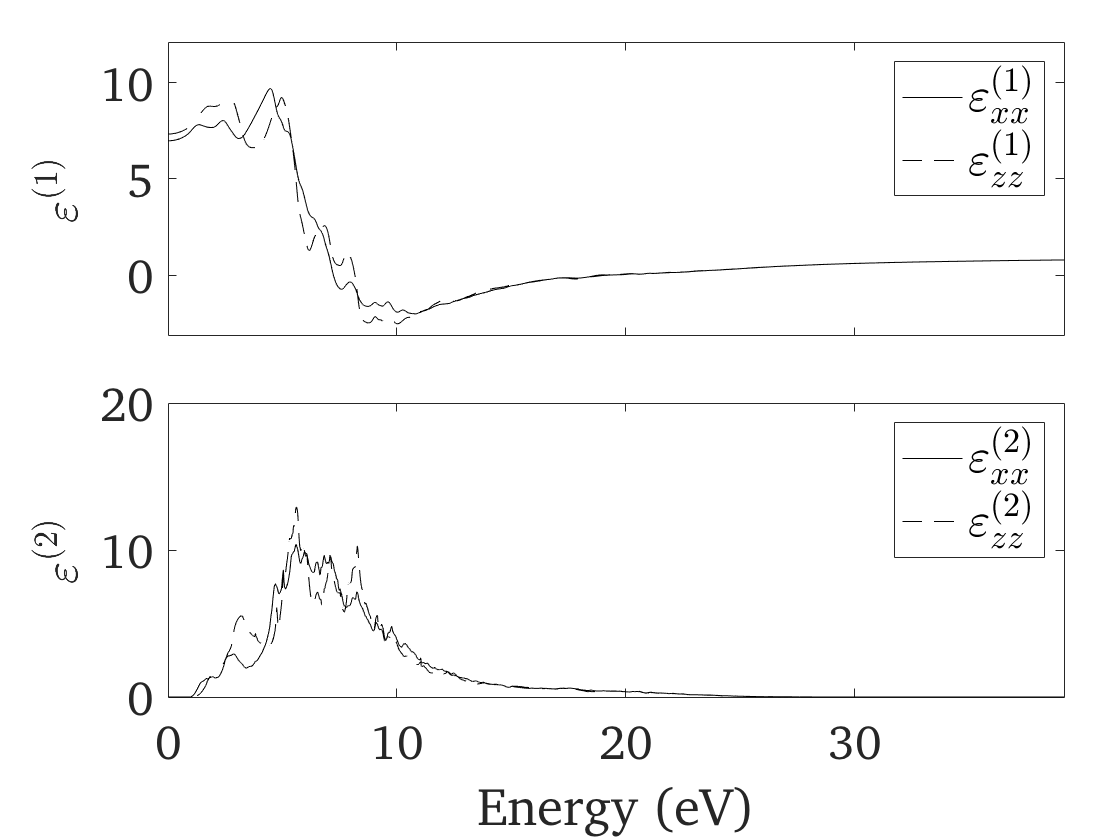}
\caption{}
\end{subfigure}
\begin{subfigure}{0.4\textwidth}
\centering
\hspace{3em}CuGaTe$_2$
\includegraphics[width=0.8\linewidth]{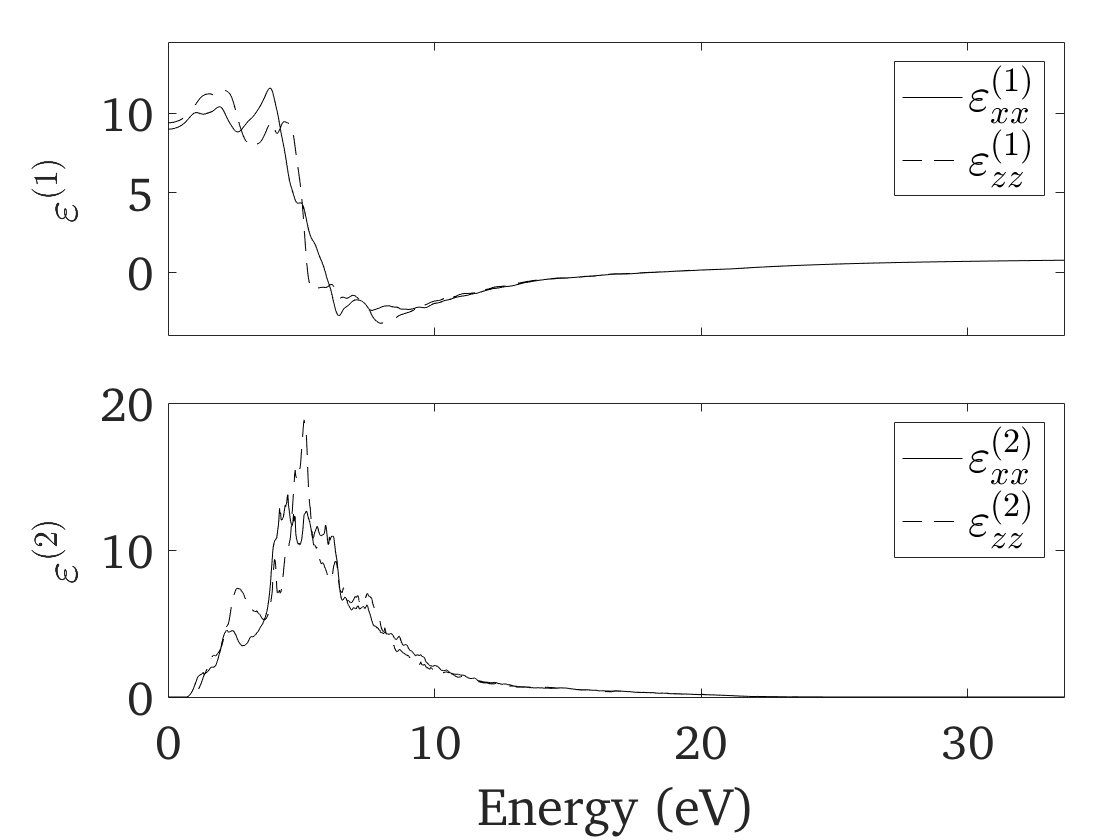}
\caption{}
\end{subfigure}%
\begin{subfigure}{0.4\textwidth}
\centering
\hspace{2em}CuInS$_2$
\includegraphics[width=0.8\linewidth]{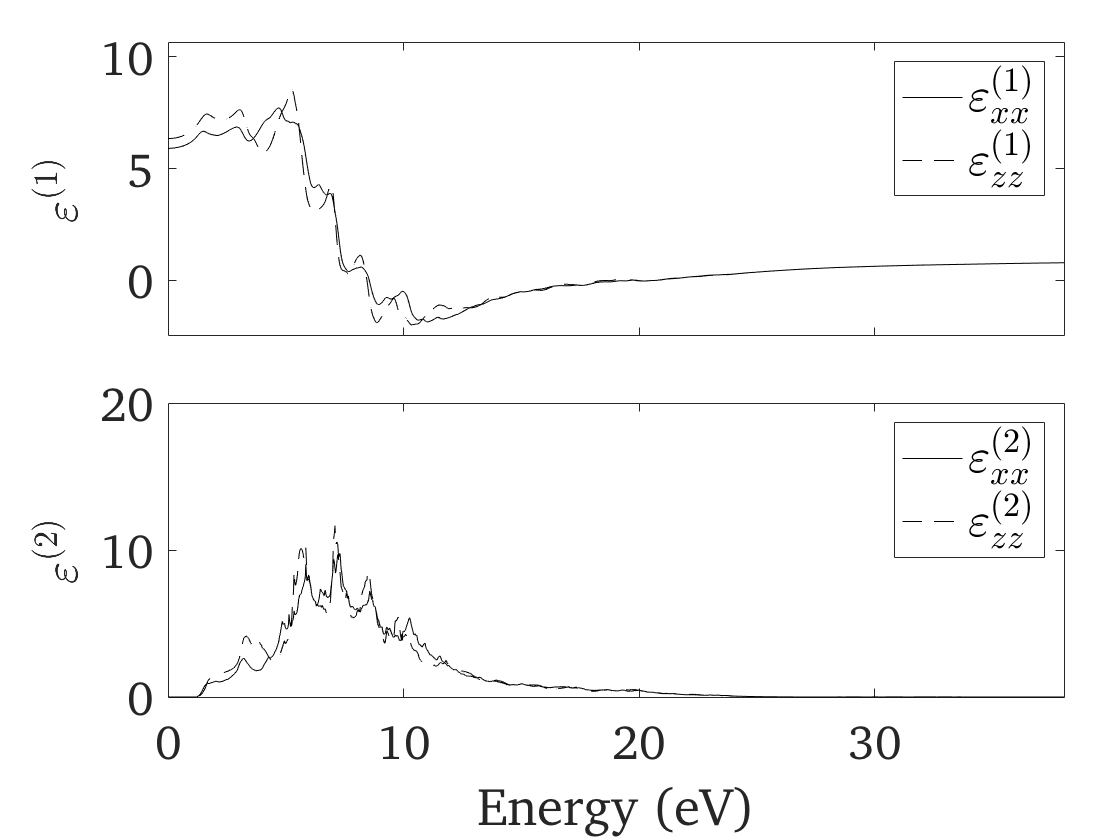}
\caption{}
\end{subfigure}
\begin{subfigure}{0.4\textwidth}
\centering
\hspace{3em}CuInSe$_2$
\includegraphics[width=0.8\linewidth]{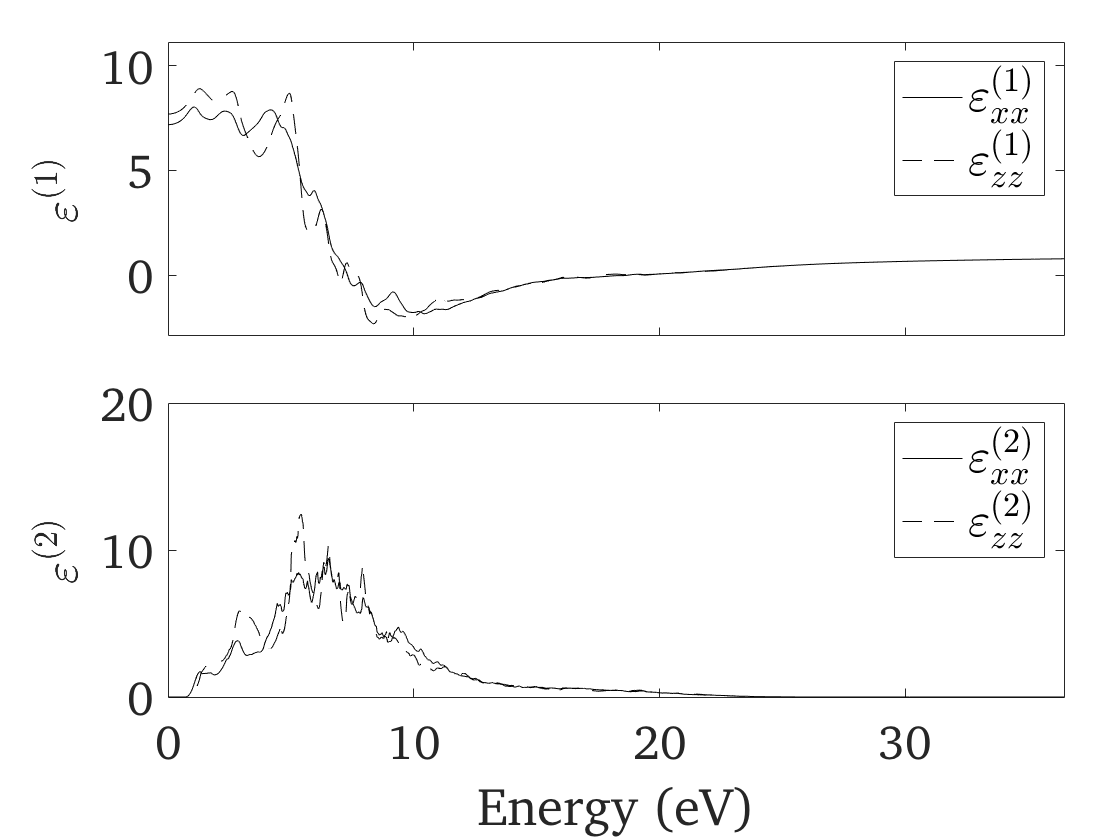}
\caption{}
\end{subfigure}%
\begin{subfigure}{0.4\textwidth}
\centering
\hspace{2em}CuInTe$_2$
\includegraphics[width=0.8\linewidth]{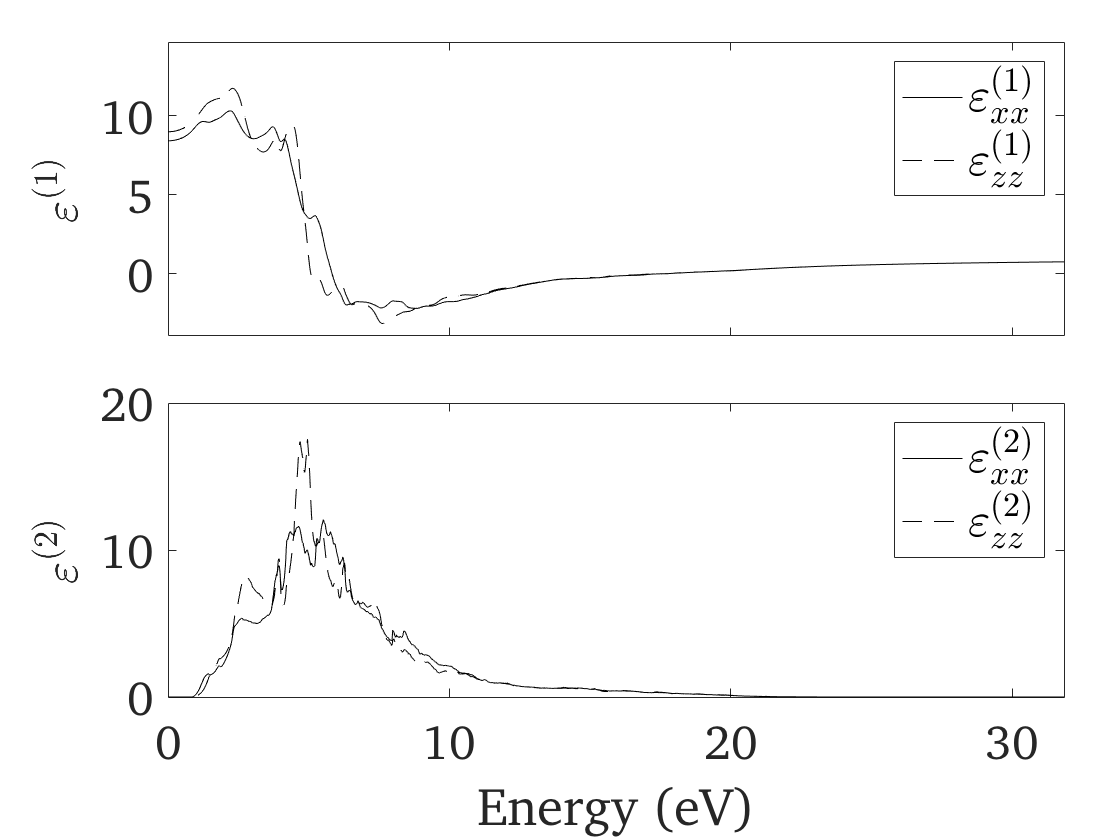}
\caption{}
\end{subfigure}
\caption{\label{fig:Cudiel}Dielectric functions of the CuAu-like phase of the Cu-III-VI$_2$ compounds.}
\end{figure}

\end{document}